# Integrative genomics analysis identifies pericentromeric regions of human chromosomes affecting patterns of inter-chromosomal interactions

**Gennadi V. Glinsky, MD, Ph.D.**

The Stanford University School of Medicine, Medical School Lab Surge Bldg, Room P214, 1201 Welch Road, Stanford, CA 94305-5494

The Sanford-Burnham Medical Research Institute, 10901 North Torrey Pines Road, La Jolla, CA 92037

The Rockefeller University, Laboratory of Immune Cell Epigenetics & Signaling, Weiss Research Building, 8[th] Floor, 1230 York Avenue, New York, NY 10065-6399

**Abstract**

Genome-wide analysis of distributions of densities of long-range interactions of human chromosomes with each other, nucleoli, nuclear lamina, and binding sites of chromatin state regulatory proteins, CTCF and STAT1, identifies non-random highly correlated patterns of density distributions along the chromosome length for all these features. Marked co-enrichments and clustering of all these interactions are detected at discrete genomic regions on selected chromosomes, which are located within pericentromeric heterochromatin and designated Centromeric Regions of Interphase Chromatin Homing (CENTRICH). CENTRICH manifest 199-716-fold higher density of inter-chromosomal binding sites compared to genome-wide or chromosomal averages (p = $2.10E^{-101}$-$1.08E^{-292}$). Sequence alignment analysis shows that CENTRICH represent unique DNA sequences of 3.9 to 22.4 Kb in size which are: 1) associated with nucleolus; 2) exhibit highly diverse set of DNA-bound chromatin state regulators, including marked enrichment of CTCF and STAT1 binding sites; 3) bind multiple intergenic disease-associated genomic loci (IDAGL) with documented long-range enhancer activities and established links to increased risk of developing epithelial malignancies and other common human disorders. Using distances of SNP loci homing sites within genomic coordinates of CENTRICH as a proxy of likelihood of disease-linked SNP loci binding to CENTRICH, we demonstrate statistically significant correlations between the probability of SNP loci binding to CENTRICH and GWAS-defined odds ratios of increased risk of a disease for cancer, coronary artery disease, and type 2 diabetes. Our analysis suggests that centromeric sequences and pericentromeric heterochromatin may play an important role in human cells beyond the critical functions in chromosome segregation.

**Introduction**

In human genome many gene deserts are significantly enriched for predicted long-range enhancers and/or insulators genomic coordinates of which are in close proximity or overlap with disease-associated SNPs (1-3). Several groups experimentally validated the biological activity of intergenic long-range enhancers and reported active transcription at intergenic disease-associated genetic loci (IDAGL) with documented increased risk of developing prostate cancer (3-8), autoimmune disorders (1, 3), coronary artery diseases and type 2 diabetes (2). However, experimental observations and theoretical considerations highlighting detailed mechanisms of structural-functional alignment and compartmentalization of genomic functions in interphase nuclei with respect to regulatory elements within gene deserts are lacking.

In interphase nucleus, the three-dimensional conformation of chromosomes is defined by both chromatin folding and chromatin polymer interactions with major external to chromatin globules nuclear structures such as nuclear lamina and nucleoli (9-13). These interactions are likely to influence the compartmentalization of the nuclear space into functionally-related domains which brings widely separated genetic loci into close, spatially-defined proximity. Therefore, deeper mechanistic understanding how chromosomes align within the three-dimensional space of interphase nucleus can provide important insight into the complex relationships between chromatin architecture, regulation of gene expression, and functional states of a cell.

Recently, a genome-wide, high-resolution analysis of DNA sequences that have a high probability to associate endogenously and co-purify with nucleoli (nucleolar-associated domains [NADs]) in human cells, has been carried-out using a combination of fluorescence comparative genome hybridization (CGH), deep DNA sequencing and fluorescence photoactivation (9, 10). It has been noted that the size distribution (0.1–10 Mb) and median sequence length (749 kb) of NADs were very similar to lamina-associated domains, LADs (0.1–10 Mb, 553 kb; ref. 11) suggesting that the major architectural units of chromosome organization within the mammalian interphase nucleus are about 0.5–1 Mb in length (9). Genome-wide analysis of interactions of interphase chromatin with nuclear lamina in mouse embryonic stem cells revealed that nuclear lamina-genome interactions are broadly involved in the control of gene expression programs during lineage commitment and terminal differentiation of embryonic stem cells (12). Therefore, recent experiments on large cell populations demonstrate that interactions of interphase chromosomes with nuclear lamina and nucleoli are critically important in defining a 3D architecture of chromatin organization and functions. Expression of genes located in both lamina-associated and nucleoli-associated chromatin compartments is predominantly repressed (9-12). Conversely, transcription at genetic loci that are located in chromatin compartments that are not associated with either nuclear lamina or nucleoli is often activated and expression of corresponding genes is increased.

Experimental studies of long-range physical interactions of interphase chromatin chains utilize multiple methods and have evolved from analysis of interactions between specific pairs of distant genomic loci to genome-wide approaches. Chromosome Conformation Capture (3C) utilizes spatially constrained ligation followed by locus-specific PCR. Adaptations and modifications of 3C have extended the analytical process with the use of inverse PCR (4C) or multiplexed ligation-mediated amplification (5C). A genome-wide approach to the analysis of long-range chromatin interactions named Hi-C was reported that extends the above methods to enable purification of ligation products followed by massively parallel sequencing (13). Hi-C experiments allowed unbiased identification of chromatin interactions across an entire genome and provided strong support for the fractal globule model of interphase chromatin folding. Continuous flow of exceedingly large volumes of data from genome-wide chromatin state mapping experiments using Chip-seq, Chip-exo, ChIA-PET methods, investigations of replication timing profiles, and functional studies of GWAS-defined intergenic disease-associated genomic loci (1-8; 14-19) dictates the necessity of integrative genomics analyses of human interphase chromosomes.

    Here we report the results of the external points of contact analysis of chromatin chains based on a genome-wide alignment of inter- and intra-chromosomal interactions within the context of interphase chromatin binding to nuclear lamina, nucleolus, and chromatin state regulatory proteins CTCF and STAT1. We analyzed distributions of densities of interphase chromosome interactions which were uniquely mapped to the reference human genome by quantifying the number of chromosome binding events within consecutive segments of 1 Mb and 1 Kb in length of each human chromosome. We reasoned that this approach would estimate the density of interphase chromatin chain binding to a given segment of chromosomes and may highlight the genomic regions with distinct ability to attract the binding of interphase chromatin chains. This analysis may help to identify discrete chromosomal regions with increased inter-chromosomal interaction density and reveal the potentially biologically important chromosomal segments with a common function to attract binding of interphase chromatin chains of multiple (perhaps, all) human chromosomes and, therefore, could be defined as chromosomal regions of interphase chromatin homing. Consistent with this hypothesis, our analysis identifies near-centromeric regions of 3.9 – 22.4 Kb in size on selected human chromosomes which represent homologous sequences of intergenic DNA defined by the following common characteristics: 1) association with nucleolus; 2) marked enrichment for interphase chromatin homing sites; 3) remarkably diverse set of DNA-bound chromatin regulatory proteins, including enrichment of CTCF and STAT1 binding sites; 4) binding of multiple intergenic disease-associated genomic loci (IDAGL) with documented long-range enhancer activities and established links to increased risk of

developing epithelial malignancies and other common human disorders. We propose to define these regions with increased inter-chromosomal interaction density as CENTRICH and postulate that CENTRICH may play an important role in 3D genome architecture and function by governing a dynamic transition of large segments of interphase chromosomes from the open loop to folded chromatin chain conformations.

## Results

**Genome-wide maps of the interphase chromatin binding to nuclear lamina and Hi-C-defined chromatin folding compartments reveal highly correlated profiles inversely aligned along human chromosomes.**

Plotting the chromatin compartments defined by various methods linearly along the length of each chromosome revealed the intermitting chromosome-specific patterns of regions with positive and negative values (Figure 1). Regions with positive values reflect the enrichment in Hi-C compartment A and chromatin domains bound to nuclear lamina and/or nucleoli and regions with negative values reflect the enrichment in Hi-C compartment B and chromatin domains not associated with either nuclear lamina or nucleoli (Figure 1). Alignment of Hi-C interaction maps and binding profiles of Lamin B1 and nucleoli identified the striking similarities between independently defined genome-wide interphase chromatin patterns both in terms of the overall resemblance of patterns and in functional designations of the associated chromatin loci defined as predominantly silent heterochromatin (Hi-C compartment B; LAD and NAD domains) and active chromatin (Hi-C compartment A; chromatin loci not associated with either nuclear lamina or nucleoli). To quantify these relationships genome-wide, we correlated Hi-C model data (eigenvector values) for each chromosome with Lamin B1 and nucleoli binding data. We found that for all human chromosomes, a significant correlation exists between binding of chromosomal loci to nuclear lamina and segregation into spatially-defined distinct compartments of genome-wide chromatin interactions identified by Hi-C method (Figure 1; Supplemental Table S1; Supplemental Figure 1). Therefore, loci in compartment B, which corresponds to the chromatin domains bound to the nuclear lamina, exhibit a stronger tendency for close spatial localization and represent close, less accessible chromatin state, whereas loci in compartment A, which corresponds to the chromatin domains not associated with the nuclear lamina, are more closely associated with open, accessible, actively transcribed chromatin. This interpretation is in a good agreement with the observations that loci in compartment B showed a consistently higher interaction frequency at a given genomic distance than pairs of loci in compartment A indicating that compartment B is more densely packed and FISH data are consistent with this observation (13).

Careful analysis of deep sequencing data by van Koningsbruggen et al. (9) detected a significant overlap between the NADs and loci previously reported to associate with the nuclear envelope (LADs; ref. 11). Across the genome, analysis of Lamin B1 binding and association with nucleolus consistently identifies regions of heterochromatin with overlapping genomic coordinates, including multiple transcriptionally-competent intergenic disease-associated genomic loci (IDAGL; Figure 2) which produce biologically active trans-regulatory snpRNAs (1, 3). These observations suggest that chromatin binding to Lamin B1 and association with nucleoli may impact genome-wide chromatin interactions by shuttling heterochromatin loci defined by overlapping locations and affecting dynamic features of a fractal globule. Periodic directional movements of specific regions of the chromatin fractal globules would likely to depend on relative strengths of association to nucleoli and may contribute to different availability of distinct chromatin chains for engagements in interchromosomal interactions. Consistent with this prediction, we found that there is a significant direct correlation between the relative strengths of the association to nucleoli of centromeric regions of interphase chromatin homing on human chromosomes and increased likelihood of interchromosomal versus intrachromosomal interactions (Figure 1), that is human chromosomes with higher nucleoli/genomic ratios of chromatin binding exhibit increased ratios of genome-wide interchromosomal to intrachromosomal interactions. This model is in agreement with the idea that the apparent overlap in LAD and NAD loci may indicate that specific regions could alternate between binding to the nucleolus and the nuclear lamina either in different cells, or at different times within the same cells (9).

Proposed model of 3D architecture of interphase chromatin dynamics appears highly consistent with previously discovered strong correlation between replication timing and spatial proximity of chromatin as measured by Hi-C analysis that indicates that early and late initiation of replication occurs in spatially separate nuclear compartments, but rarely within the intervening chromatin (19). As shown in Supplemental Figure S2 for chromosome 17, early replication timing chromatin compartments correspond to loci in Hi-C compartment A, which corresponds to the chromatin domains not bound to the nuclear lamina and is more closely associated with open, accessible, actively transcribed chromatins. Late replication timing chromatin compartments correspond to loci in Hi-C compartment B, which corresponds to the chromatin domains bound to the nuclear lamina and nucleolus, exhibits a stronger tendency for close spatial localization and represents less accessible, more densely packed and less actively transcribed heterochromatin state (Supplemental Figure S2).

**Identification of Centromeric Regions of Interphase Chromatin Homing (CENTRICH).**

In human genome many gene deserts are significantly enriched for predicted long-range enhancers and/or insulators genomic coordinates of which are in close proximity or overlap with disease-associated SNPs (1, 3). Several groups experimentally validated the presence of functional enhancers and transcriptional activity in intergenic disease-associated genetic loci (IDAGL) with documented increased risk of developing prostate cancer (3-8), autoimmune disorders (1, 3), coronary artery diseases and type 2 diabetes (2). We performed a genome-wide survey of Hi-C-defined long-range physical chromatin interactions utilizing chromosomal positions of 99 intergenic disease-associated SNPs as probes to identify genomic coordinates of interacting pairs of enhancers and targeted genetic loci. Unexpectedly, our analysis revealed that all IDAGL home to specific discrete near-centrometic regions on human chromosomes 2 and 10 (Figure 2). We designated these regions as the Cetromeric Regions of Interphase Chromatin Homing (CENTRICH). To determine whether CENTRICH are unique for human chromosomes 2 and 10, we performed genome-wide analysis of the frequencies (densities) of long-range interactions along the chromosome length for all human chromosomes. To this end, we computed all experimentally documented interactions along each chromosome consecutively segregated into 1Mb windows, plotted results of the analysis and look for regions that manifest statistically higher densities of long-range interactions. To evaluate statistical significance of the differences in the densities of long-range chromatin interactions, we compared the candidate CENTRICH frequency values to genome-wide frequency (test 1), individual chromosomal frequency (test 2), and interaction frequencies of neighboring regions (test 3; neighboring regions were 5 Mb segments on both sides of the candidate CENTRICH). Using these criteria, we identified ten CENTRICH on human chromosomes 2; 10; 17; 1; 7; 4; 21; 19; 16; 12 (Figures 1 & 2). For ten CENTRICH displaying statistically significant enrichments of interphase chromatin homing sites within 1 Mb windows, we defined the precise genomic coordinates of the enrichment regions by performing the follow-up analyses at the 100 Kb and 1 Kb resolution windows. Four CENTRICH on chr2; chr10; chr17; and chr1 (Table 1) represent most prominent interphase chromatin homing sites of human genome: they engaged in 397-1526 pair-wise interactions per 1 Kb distance, which represents 199 - 716-fold enrichment of interphase chromatin homing sites compared to genome-wide average (2-tail Fisher's exact test p values range $2.10E^{-101}$ - $1.08E^{-292}$). Comparisons of CENTRICH sequences reveals that CENTRICH on different chromosomes display statistically significant homology (Table 1 & Supplemental Table S2) with maximum sequence identities defined by BLAST search ranging from 76% (query coverage 99%; E value = 0; total score = 2.99E+05) to 88% (query coverage 100%; E value = 0; total score = 2.78E+05). Query coverage range in all pair-wise searches was from 99% to 100% and total score range was from 6.57E+04 to 1.43E+06 (E value = 0 in all pair-wise comparisons). Systematic BLAST searches using chr2 CENTRICH sequence as a query in a megablast algorithm setting and review of the vertebrates and mammalian conservation data for CENTRICH genomic regions (Placental Mammalian Basewise Conservation by PhyloP and Multiz Alignments of 44 Vertebrates) at the UCSC genome browser (http://genome.ucsc.edu/) did not identify significant CENTRICH homologues with similar genomic arrangements in genomes of other mammals and vertebrates. These data suggest that CENTRICH are unique for human genomic regulatory sequences.

**CENTRICH represent common homing sites for both enhancers and promoters engaged in the long-range regulatory interactions.**

Genome-wide transcriptional activity of human genome is governed by the extensive web of promoter-promoter and enhancer-promoter interactions of interphase chromatin chains with significant enrichment of enhancer-promoter interactions in defining cell type-specific transcription (14). These widespread interphase chromatin regulatory complexes provide topological basis for genome-wide transcriptional regulation of both housekeeping and cell-specific genes in human cells (14). Typically, the long-range chromatin interaction complexes contain multiple physically interacting enhancers and promoters and have genomic spans of hundreds kilobases. These complexes may span several megabases and contain hundreds of genes as complexes, which are distantly separated on a chromosome or located on different chromosomes, further converge to form high-order multigene interactions regulatory complexes. The chromatin looping model was proposed to explain how these multigene clusters were build and function. According to looping model, the enhancer comes into close spatial proximity with the target promoter guided by the protein-protein interactions of DNA-bound chromatin state regulators. This triggers transcriptional regulatory events, establishes stable enhancer-promoter complex due to action of the Mediator/Cohesin complex, and loops out the intervening chromatin chains. Precise mechanisms of the molecular assembly of these interphase chromatin looping structures and role in this process of the intervening chromatin chains remains unknown.

We thought to analyze the binding patterns to human CENTRICH of functionally-validated multigene interphase chromatin chain clusters with documented coordinated transcriptional regulation of protein-coding genes and established long-range enhancer-promoter interactions which were confirmed by several independent methods, including qPCR, 3C, and DNA FISH (Figures 3 & 4). Remarkably, in all instances we found that high-density regions of the CENTRICH binding sites to the long-range interphase chromatin complexes are located within intervening chromatin chains (Figures 3 & 4). Conversely, the genomic segments of experimentally documented long-range enhancer-promoter and promoter-promoter interactions appear excluded from binding to CENTRICH; they are not engaged in binding to CENTRICH beyond the levels of binding expected by chance. The high-density homing sites of the long-range chromatin complexes within the CENTRICH are placed in close spatial proximity to each other in topological arrangements compatible with a model of formation of neighboring looping structures on a common CENTRICH base (Figure 4). Notably, the inner genomic elements of these CENTRICH-associated neighboring looping structures which are housed on a CENTRICH base appear to include 75%-100% of experimentally documented long-range enhancer-promoter interactions within the given single-gene or multigene interphase chromatin complex (Figures 3 & 4). Taken together, these data are consistent with the model that physical interactions between CENTRICH and long-range interphase chromatin enhancer-promoter complexes occur primarily within the intervening chromatin regions. Binding of long-range interphase chromatin enhancer-promoter complexes to the CENTRICH via high-frequency interactions with intervening chromatin may promote formation of the meta-stable intermediary structures consisting of neighboring CENTRICH-associated chromatin loops which are housed on a CENTRICH base. Formation of intermediary CENTRICH-associated neighboring chromatin loops would place in close spatial proximity functionally-compatible promoter-promoter and enhancer-promoter pairs and facilitate formation of stable long-range chromatin complexes between multiple enhancers and promoters. Formation of these secondary stable chromatin complexes would involve engagement of the Mediator/Cohesin complex; trigger conformational changes and dissociation of intervening chromatin chains from the CENTRICH resulting in looping structures of intervening chromatin. Consistent with this model, CTCF and Rad21, two key regulatory proteins which are known to play essential roles in long-range enhancer-promoter interactions during targeted activation of transcription of protein-coding genes, are members of the elite consensus set of chromatin state regulatory proteins which are consistently bound to DNA within human CENTRICH.

We carried out similar analysis for 99 IDAGLs and found that many protein-coding genes that were experimentally validated as regulatory targets of long-range enhancers and/or snpRNAs have multiple homing sites within the chr2 and chr10 CENTRICH (Figures 2 & 3; Supplemental Figures S4-S6; unpublished observations). These data indicate that both long-range enhancers and regulatory target loci are bound to the chr2 and chr10 CENTRICH. Binding of both long-range enhancers and regulatory target loci to the CENTRICH would place them in close spatial proximity within the nuclear space and facilitate physical interactions and clustering thus increasing the probability of phenotypically relevant regulatory events. Consistent with this model, for cancer, coronary artery disease, and type 2 diabetes, there is a statistically significant inverse correlation between the genome-wide association studies (GWAS)-defined odds ratios of increased risk of a disease and distances of SNP loci homing sites to the middle-point genomic coordinates of CENTRICH on chromosome 2 (Figure 2 and Supplemental Figures S4-S6). Considering distances of SNP loci homing sites within genomic coordinates of CENTRICH as a proxy of likelihood of SNP loci binding to CENTRIH, this analysis implies a direct correlation between higher probability of disease SNP loci homing within CENTRICH, which corresponds to smaller distances of SNP loci homing sites in Figures 2 and Supplemental Figures S4-S6, and increased likelihood of developing a disease, which corresponds to higher values of GWAS-defined odds ratios of increased risk of a disease in Figures 2 and Supplemental Figures S4-S6.

**Sequence alignment analysis of human CENTRICH.**

We performed a systematic analysis of DNA sequences of human CENTRICH using multiple feature settings of the UCSC genome browser tracking windows. This analysis revealed that identified CENTRICH regions contain either alpha satellite DNA, or HSATII repeat sequences, both of which are extremely repetitive in the genome. It poses questions whether the CENTRICH are unique genomic sequences that could be reliably mapped to a single genomic location in the human genome and raises concerns regarding the reliability of the sequence alignments and mapping to the defined genomic regions of the original Hi-C data. Some of these concerns are based in part on the a priori assumptions that genomic sequences containing arrays of highly repetitive elements cannot be reliably mapped because they are not uniquely represented in the human genome. Acceptance of this notion and filtering out all such regions without detailed experimental analyses may cause loss of potentially significant biological signals (false negatives). We found that this assumption is not true for many regulatory and disease-associated sequences in gene deserts, intronic, and intergenic regions of human genome. Despite the fact that many regulatory sequences in gene deserts, intronic, and intergenic regions contain clearly identifiable arrays of sequences homologous to repeats; they represent unique genomic sequences in the reference human genome. These data indicate that in many instances the arrays of highly repetitive elements, particularly satellite repeats in centromeric regions, placed consecutively in distinct parts of human genome evolved into unique genomic sequences. Best known examples of arrays of satellite repeats forming unique genomic sequences are higher order repeats of human centrmeric regions on each chromosome. Therefore, a priori assumption that the alignment and mapping of sequencing data cannot be reliably performed within the genomic regions containing repetitive elements is not correct and the validity of this assumption should be always experimentally tested.

Several independent lines of experimental evidence invalidate the arguments of potential sequence alignment and mapping artifacts within the CENTRICH. Review of the genomic mapping data from multiple independent groups of investigators reported in the UCSC human genome browser demonstrates that thousands of experimentally-defined genomic features were placed at defined positions within the CENTRICH sequences. These include among many other features the uniquely mapped SNP variants from the dbSNP database and HapMap consortium, histone variants, transcription factors, chromatin maintenance proteins, short and long non-coding RNA sequences from the ENCODE projects, and many other data points and experimental variables which are readily available for follow-up analyses and visualization using the corresponding tracking windows and tools of the UCSC Genome Browser (http://genome.ucsc.edu/ ). Large number of well-documented peer-reviewed studies, which represent a cornerstone of the USC Human Genome Browser database and reported successful mapping of thousands of distinct genomic features at the unique positions within the CENTRICH, were executed using not only multiple cultured human cell lines, but also primary human cells, and many different human tissues from different individuals. This is an important fact because it makes unlikely the potential mapping artifacts arising from comparisons of genomes of cultured cell lines and DNA sequences derived from primary (non-cultured) human cells and tissues within the context of the reference human genome.

According to original publications, analyzed in this study sequencing read tags were generated and uniquely aligned to the reference human genome using state of the art deep sequencing technologies and analytical tools with the level of confidence deemed sufficient to define a match at a single location in the reference human genome (9-18). Nevertheless, we thought to address the potential problem of artifacts in mapping to the CENTRICH sequences of the original experimental sequencing data by performing several additional analyses. The reliability and accuracy of the reference human genome sequencing data is extremely high with less than one error per 10,000 bases (http://genome.ucsc.edu/cgi-bin/hgGateway ). Therefore, the human genome reference sequence is considered highly accurate and contiguous with the only remaining gaps in the regions whose sequences cannot be resolved using current technology. There are no unresolved sequence gaps within the CENTRICH. We considered in our analyses as unique genomic sequences the query sequences that have a single identical counterpart in the reference human genome defined by 100% maximal identity with 100% query coverage ad e-value 0 using the MEGABLAST algorithm. All CENTRICH regions reported in the Table 1 satisfy this stringent definition of unique DNA sequences in human reference genome database. We carried out the systematic review of the original alignment and mapping data and determined that DNA sequences of each CENTRICH listed in the Table 1 were defined by the essentially continuous linear alignments of thousands of experimental sequence reads derived from biological replicate experiments. We thought to test whether the definition of unique genomic sequences is true for DNA sequences within the CENTRICH regions by dividing each CENTRICH into series of short continuous segments along the entire length of the region and covering the CENTRICH sequences essentially without any unresolved sequence gaps (Table 2).

We utilized the MEGABLAST algorithm to perform the sequences alignment analyses for randomly generated short DNA segments of the CENTRICH regions. We found that short continuous DNA segments within the CENTRICH represent unique sequences in human reference genome database. Results of this type of analysis are reported in the Table 2 for 29 short continuous DNA segments within the CENTRICH located on four different chromosomes. Based on cumulative evidence generated by these analyses we conclude that potential sequence alignment and genomic mapping artifacts are not likely source of experimental and analytical errors in definition of genomic coordinates of human CENTRICH.

**CENTRICH are highly enriched for CTCF and STAT1 binding sites.**

We thought to use the alternative analytical means to the interrogation of Hi-C data for identification of distinct genomic features of human CENTRICH. Survey of the ENCODE chromatin state maps shows that CENTRICH display highly diverse panels of transcription factors and chromatin state regulatory proteins bound to DNA at the attachment sites of IDAGL (see above). CTCF and STAT1 are two members of the elite consensus set of master chromatin state regulators bound to genomic DNA within all identified CENTRICH which were directly implicated as key regulatory proteins playing significant roles in chromatin assembly and functions.

    Human CTCF binding data deserves particular attention because this data set was generated using novel Chip-exo method for comprehensive genome-wide mapping of protein-DNA interactions at single-nucleotide resolution (15). Chip-exo was specifically designed and validated to address limitations of the ChIP-chip and ChIP-seq assays producing false-positives calls which are associated with the presence of contaminating DNA and DNA fragmentation heterogeneity in the chromatin immunoprecipitation samples subjected to deep sequencing. It also appears to address the problem

of missed calls (false negatives) which produced by the stringent data filtering (15). Chip-exo detected 35,161 CTCF-bound genomic locations in human cells, which confirmed coordinates of 93% of CTCF-bound regions determined previously by ChIP-seq (20) and revealed another 17,000 locations that were missed (15).

Analysis of the genome-wide database of CTCF binding sites generated by the Chip-exo method demonstrates that human CENTRICH are highly enriched for CTCF-bound locations (Figure 5 and Table 3). In many instances Chip-exo method identifies CENTRICH as the prominent genomic locations which are markedly enriched for high-density CTCF binding sites which indicate that increased distribution frequencies of CTCF-binging sites could have been used as alternative analytical means for identification of these genomic regions.

We noted that CENTRICH on different chromosomes manifest significantly distinct levels of enrichment of interphase chromatin binding sites (Table 1). Correlation analysis demonstrates that levels of enrichments for interphase chromatin homing sites and high-density CTCF-binding sites on different CENTRICH are significantly correlated (Figure 5). These observations indicate that high density of the CTCF-binding sites may contribute to the increased propensity of CENTRICH to attract binding of interphase chromatin chains.

Similar to the CTCF-binding sites, analysis of the genome-wide database of uniquely mapped STAT1 binding sites (16) demonstrates that human CENTRICH are significantly enriched for high-density STAT1-bound locations (Figure 5 and Table 4). Notably, genomic locations of high-density CTCF- and STAT1-binding sites display highly correlated profiles in human genome (Figure 5). In contrast to other genomic locations, the level of enrichment and distribution within CENTRICH of the high-density STAT1-binding sites appear not altered in response to interferon treatment (Figure 5 and Table 4), suggesting distinct mechanisms of chromatin state regulation in these regions. Based on these analyses, we conclude that CENTRICH are significantly enriched for both CTCF- and STAT1-binding sites.

Discussion

**Interactions with nuclear lamina and nucleoli govern dynamic features of the chromatin chains.** Application of the Hi-C method to the analysis of interphase chromatin folding revealed the existence of two independent compartments of genome-wide chromatin interactions along the chains of chromatin polymers which were designated compartment A and compartment B (13). The compartments were defined solely based on Hi-C data primarily taking into account the enrichment of contacts between DNA sequences within each compartment and depletion of contacts between sequences in different compartments. Compartment B showed a higher interaction frequency, lower overall DNaseI accessibility, and lower gene expression, indicating that compartment B represents more densely packed heterochromatin. In contrast, compartment A showed a relatively lower interaction frequency, higher overall DNaseI accessibility, and increased gene expression, indicating that compartment A represents less densely packed, transcriptionally active chromatin. The Hi-C-defined chromatin compartments corresponded to spatially separated regions of chromosomes as confirmed by fluorescence in situ hybridization (FISH). Plotting of these compartments linearly along the length of each chromosome using the eigenvector index values revealed the intermitting chromosome-specific patterns with enrichment in compartment A as positive values and enrichment in compartment B as negative values (13). We noted the striking similarities between independently defined genome-wide patterns of Hi-C interaction maps and lamin B1 and nucleoli binding profiles both in terms of the overall resemblance of patterns and in functional designations of the associated chromatin loci defined as predominantly silent heterochromatin (Hi-C compartment B; LAD and NAD domains) and active chromatin (Hi-C compartment A; chromatin loci not associated with nuclear lamina and/or nucleoli). These preliminary observations suggest that interphase chromatin compartments which were independently defined using distinct methodological approaches may be structurally and, perhaps, mechanistically and functionally related. However, analysis of the fractal globule model of interphase chromatin geometry was lacking systematic considerations of the impact of nuclear matrix and nucleolus.

Here we performed genome-wide alignment of Hi-C interaction maps and binding profiles of Lamin B1 and nucleoli to determine whether the apparent descriptive similarities can be formally quantified and assigned to specific genomic coordinates. Our analyses reveal highly significant genome-wide correlations between independently defined interphase chromatin chain binding to Lamin B1 and nucleoli and fractal's folding patterns both in terms of the assignments of specific genomic coordinates and in functional designations of the associated chromosomal regions defined as predominantly silent heterochromatin (Hi-C compartment B; LAD and NAD domains) and active chromatin (Hi-C compartment A; chromatin loci not associated with either nuclear lamina or nucleoli).

We quantified these relationships for all human chromosomes by calculating Hi-C model data (eigenvector values), Lamin B1, and nucleoli binding data along the length of each chromosome segregated into discrete segments of 1 Mb, 100 Kb, and 1 Kb. We reasoned that this approach should highlight the distribution of densities of interphase chromatin chain interactions and corresponding binding sites on different chromosomes and within distinct chromosomal regions. We found that for all human chromosomes, a significant correlation exists between genomic locations of chromosomal regions bound to nuclear lamina and chromosomal segments segregated into compartment B by the Hi-C method (Figure 1; Supplemental Table S1; Supplemental Figure 1). We conclude that genomic loci in the Hi-C compartment B correspond to the chromatin domains bound to the nuclear lamina, exhibit a stronger tendency for close spatial localization and represent close, less accessible chromatin state. Conversely, genomic loci in the Hi-C compartment A, which corresponds to the chromatin domains not associated with the nuclear lamina, are more closely associated with open, accessible, actively transcribed chromatin. Formal alignment of the lamina-associated genomic regions and Hi-C compartment B is in a good agreement with the observations that loci in the compartment B showed a consistently higher interaction frequency at a given genomic distance than pairs of loci in the compartment A (13).

Taken together, these observations and FISH data (13) are consistent with definition of the Hi-C compartment B as chromosomal domains containing densely packed interphase chromatin chains bound to either nuclear lamina or nucleoli. Mechanistically, our analysis implies that Hi-C-defined chromatin compartments A and B are not reflecting intrinsic features of interphase chromatin chains. They are likely to reflect the interactions of chromatin chains with nuclear lamina and nucleoli, indicating that binding to nuclear lamina and nucleoli may play an important role in defining folding patterns of interphase chromatin chains and global regulation of inter- and intrachromosomal interactions in interphase nucleus. Similarly, interactions of chromatin chains with nuclear lamina and nucleolus may have a major regulatory impact on definition of the early and late replication timing compartments (20).

**Heterochromatin "shuttle" model of the chromatin chain's dynamics.** Across the genome, applications of three independent methods of genome-wide interphase chromatin analysis (Hi-C chromosome conformational capture, Lamin B1 binding, and association with nucleolus) consistently identify regions of heterochromatin with overlapping genomic coordinates, including multiple loci harboring disease-linked SNPs. These observations suggest that chromatin binding to Lamin B1 and association with nucleoli may impact genome-wide chromatin interactions by shuttling overlapping heterochromatin loci and affecting dynamic features of a chromatin globule. According to this model, chromatin binding to either nuclear lamina or nucleolus initiates a polymer collapse and folding, promotes spatial segregation of genomic and chromosome territories, and contributes to relative stability of a fractal globule conformation (Figure 6). Shuttling between nuclear lamina and nucleoli would facilitate chromatin unfolding and loop opening, which would enable a genome-wide cross-talk due to chromatin loop invasion within and across territories and enhance the probability of long-range inter- and intra-chromosomal interactions (Figure 6). A fractal globule conformation would allow a rapid chromatin loop expansion (13, 21), crossing of boundaries of genomic and chromosome territories, enhancing a likelihood of long-range enhancer/promoter interactions and contributing to formation of transcription factories.

One of the interesting features of this model is the prediction that biomechanical forces created during a two-fold expansion of the nuclear volume in G1, changes of nuclear shape, and periodic nuclear rotations would impact the equilibrium of chromatin binding to nuclear lamina and nucleoli and influence initiation, duration, and frequency of a chromatin chain loop folding/opening cycles. It follows that dynamic transition cycles of the fractals in the nucleus may be affected by interactions of long-range intergenic enhancers with active promoters, concomitant activation of transcription at the enhancer's loci, and effects of enhancer's RNAs on the equilibrium state of fractal globule binding to the external nuclear structures. Systemic effects on fractal's dynamics of biomechanical forces during the interphase such as expansion of nuclear volume, nuclear rotation, fragmentation and assembly of nucleoli, periodic directional movements of sub-nuclear structures are likely to contribute to remarkable diversity of transcriptional landscapes and phenotypic features within populations of human cells.

Because both the LAD and NAD profiles were generated using the high-throughput analysis of large cell populations, the apparent overlap in loci may indicate that specific regions could alternate between binding to the nucleolus and the nuclear lamina either in different cells, or at different times within the same cells (9). Alternatively to the interphase heterochromatin shuttle model, the assignments of heterochromatin loci to LAD and NAD within interphase cells could remain constant following exit from mitosis and random heterochromatin placement reprogramming in daughter cells as the consequence of the nucleoli disassembly and re-assembly cycles during the cell division. Definition of genomic coordinates and precise knowledge of DNA sequences of CENTRICH provides essential information which would facilitate the conception, design, and execution of the experiments using a single cell time-lapse live video monitoring technologies to distinguish between the "heterochromatin shuttles" and "static reprogramming" models of heterochromatin maintenance and dynamics in interphase cells. Similar considerations are valid for most of the 3D chromatin conformations because they were seen in a large population of cells and their intracellular dynamics will not be refined until single-cell methods of the 3D architecture of interphase chromatin are developed and experimentally implemented.

**Mechanisms of CENTRICH interactions with the tissue-specific developmentally-regulated single-gene enhancer/promoter complexes and multigene complexes.** Our analysis appears to indicate that as a general rule the likelihood of the homing to the CENTRICH is directly correlated with the size of the genomic loci which imply that large protein-coding genes are more likely to bind to CENTRICH than short, intron-poor or intronless genes (Supplemental Figure S6 and data not shown). This mechanism may provide simple evolutionary means to use the trial and error technique in search for biological optimization of a circuitry of genetic networks by increasing gene size and placing genetic loci in the CENTRICH interactivity web. Recent experiments indicate that gene-size-

associated regulatory mechanisms distinguish developmentally-regulated or tissue-specific genes and housekeeping genes. Developmentally-regulated or tissue-specific genes share several common genomic features distinct from housekeeping genes: they are located in regions of less gene density, they have longer gene body, higher intron/exon ratio, and higher noncoding to coding region ratio than housekeeping or basal promoter genes (14; 22, 23).

It has been demonstrated that in human genome many thousands of promoter-enhancer and promoter-promoter interactions cluster multiple genes which are located at very large genomic distances often exceeding 500 kb and 1 Mb (14). These highly complex long-range chromatin looping structures markedly influence transcription in human cells (14). RNAPII-associated chromatin models classify human protein-coding genes into basal promoters, single-gene and multigene complexes based on distinct structural features, transcription activity, and chromatin state defined by epigenetic marks. Li et al. (14) reported that genes in single-gene complexes with enhancer-promoter connectivity are predominantly tissue-specific, which is in agreement with growing evidence that the expression levels of developmental and tissue-specific genes are modulated through similar mechanisms: cis-remote regulatory elements and trans-acting chromatin-associated proteins (24, 25). Protein-coding genes in the multigene complex regions were relatively shorter than other gene categories, which is a known property of highly expressed genes (22). Genes engaged in the single-gene complexes were significantly longer and had higher intron/exon ratios than the genes of other RNAPII-associated chromatin models (14), which indicates that genes with enhancer-promoter interactions in single-gene complexes were more likely to be tissue-specific or developmentally regulated, in accord with the earlier observations (22, 23). Protein-coding genes engaged in multigene complexes as well as the basal promoter genes were shown to manifest both tissue-specific and housekeeping expression patterns (14) and genomic domains of highly transcribed genes ("ridges") contain both housekeeping and tissue-specific genes (26). Summarizing these data, Li et al. (14) proposed that promoter-promoter and promoter-enhancer interactions could serve as a dominant topological framework for transcription regulation of both housekeeping and tissue-specific genes in mammalian genomes.

Our analysis demonstrates that protein-coding genes and enhancers comprising both tissue-specific developmentally-regulated single-gene enhancer/promoter complexes and multigene enhancer/promoter complexes have multiple homing sites within the CENTRICH on the same chromosome (Figures 3 & 4). These common homing sites within the CENTRICH are placed in close spatial proximity to each other in topological arrangements compatible with a model of formation of neighboring looping structures on a common CENTRICH base (Figures 3 & 4). Notably, the high density contact sites with the CENTRICH base appear placed almost exclusively within the intervening chromatin and inner genomic elements of the CENTRICH-associated neighboring looping structures appear to include 68%-100% of experimentally documented long-range enhancer-promoter interactions within the given single-gene or multigene complex (Figures 3 & 4). These data are consistent with the model that physical interactions between CENTRICH and genomic elements of the long-range enhancer-promoter complexes, which occur primarily within the intervening chromatin loops of interacting regions, may contribute mechanistically to formation of long-range enhancer/promoter looping structures, comprising core structural elements of coordinated transcriptional regulation in human genome (14).

**Limitations and shortcomings of current genome-wide chromatin state databases.** A major limitation of our analysis is defined by the fact that there is no single cell line model to which all of the essential methods of the genome-wide 3D chromatin structure analyses were applied. Despite this limitation, integration of these data derived from different cellular models within the context of the reference human genome seems to facilitate definition of novel regulatory features and networks of potential physiological and pathological relevance. Further evolution of this approach to the systems biology analysis would aim to incorporate all available information on 3D architecture of interphase chromosomes: chromatin chain folding, chromatin state maps, nuclear envelop, nucleoli, sub-nuclear structures, and nucleus, which are empirically established to be critically-important mechanistic regulatory elements in human cells. It also can be interpreted as a new dimension of the systems biology analyses which takes into account the entire genome-wide transcriptome space (in most existing models, transcriptome considerations are not complete and largely ignore gene deserts, non-protein-coding transcripts; intergenic long-range enhancers, etc.), chromatin state maps, DNA space (array CGH data; copy number, INDELs, and SNP variations), and dynamic behavior of linear polymers of interphase chromosomes folded into fractal globules. Validity of all these data and mechanistic models derived from high-throughput analyses of large cell populations must be verified using single cell methods and integrated to create a 3D model of unique regulatory architecture of interphase chromatin in individual cells. Transition to a single cell analysis level of this type of models and approaches would facilitate a deeper understanding of how continuous flow of genetic information through multidimensional matrix of deterministic and stochastic regulatory events and high-complexity web of structural elements is translated into predictable phenotypic behavior of individual cells. It would enable insights into mechanisms of integration of stochastically-defined phenotypic diversity of individual cells within populations of cells and translation of this diversity-driven heterogeneity into highly predictable and apparently programmed synergistic functional behavior of cells' populations. One of the limitations of our work is derived from the fact that original genome-wide analyses of various features of interphase chromatin chains were performed in cultured in vitro immortal human cells using different cell lines. In contrast, human genome reference data base sequences are based on DNA extracted from non-cultured human cells. If adaptation of human cells to extended culture in vitro is associated with marked expansion of CENTRICH sequences, our estimates of the average numbers of interphase homing sites could be erroneously high because they are based on calculations using the CENTRICH size estimates from human genome reference database. However, to render the enrichment of chromatin homing sites within CENTRICH statistically insignificant, the expansion of these DNA segments in cultured cells should be several hundred folds. This type of systematic large expansion of genome size in cultured human cells is particularly not likely in cells with stable diploid karyotypes. To alleviate this potential concern, we analyzed chromatin homing data from two biological replicate experiments performed on diploid GM06990 cells and reported these results in the paper. We confirmed these findings based on the analysis of all 59,309,839 alignment records of individual interactions derived from eight experiments.

We noted that in some instances on several human chromosomes the alignment profiles of distinct features of interphase chromatin chain folding manifest visible departure from the predominant correlation patterns. These could be true findings reflecting specific features of these chromosomal sub-regions or the consequence of the limitations of the currently available genome-wide data sets which were generated using different cell lines in different laboratories. We hope that our analysis and particularly the initial evidence of its potential relevance to multiple common human disorders would support arguments in favor of the critical need for continuing these types of genome-wide high-through-put experiments. It underscores the requirements for coordination of execution of these experiments between different groups to generate highly compatible multi-feature genome-wide data sets based on the same experimental models. This is particularly important now when severe shortages of research funds may diminish the enthusiasm of reviewers and funding agencies toward highly technically demanding and expensive genome-wide experiments requiring collaboration of multiple laboratories.

**Concluding remarks.** Identified in this study CENTRICH exhibit remarkably diverse regulatory context of chromatin state maps, appear engaged in self-folding, intrachromosomal and interchromosomal interactions, and attract binding of multiple IDAGL with established association to increased risk of developing epithelial malignancies and other common human disorders (Supplemental Figures S3-S5). This high-complexity molecular assortment of CENTRICH structural features seems uniquely suitable to facilitate the extraordinary high density of long-range chromatin interactions. Marked enrichment for both CTCF- and STAT1-bunding sites may play a key role in defining these unique molecular features of CENTRICH. Mechanistically related hypothesis regarding the potential role of the CTCF in regulation of the nonrandom patterns of long-range inter- and intrachromosomal interactions was proposed based on independent reanalysis of the CTCF Chip-seq and Hi-C data sets (27). Based on the analysis of the protein compositions of chromatin state maps of the CENTRICH, documented biological activities of transcription factors and chromatin remodeling proteins bound to CENTRICH, and experimental evidence of homing to CENTRICH of intergenic enhancers and protein coding genes with common patterns of transcriptional regulation, we propose several experimentally tractable hypotheses on biological functions of CENTRICH (Supplemental Figure S6).

Recent experimental evidence and theoretical considerations were generally interpreted to support the validity of the fractal globule model of a 3D architecture of interphase chromatin (13, 21, 28, 29). Our analysis reveals novel structural-functional features of interphase chromatin chains which may have a significant impact on understanding of fundamental principles of dynamic behavior of chromatin chains folding into fractal and non-fractal patterns and is an agreement with the model considering the fractal globule state as one of many transient conformations of interphase chromatin (32). Our analysis provides initial evidence that different CENTRICH manifest certain levels of specificity with respect to the binding of regulatory proteins, chromatin chains of interphase human chromosomes, and protein-coding genes (Figure 2, Supplemental Figures S3-S6, and data not shown). Strikingly, chromatin of human mitochondrion manifest 682-fold and 907-fold greater propensity to interact with CENTRICH on chrr10 and chr2, respectively, compared to other human chromosomes (Supplemental Figure S6). This may have a biological significance of safeguarding "back-up" copies of mitochondrial DNA from mutational damage in the relatively more protected nucleus compared to high exposure of the mitochondrial genome to free radicals in the mitochondrion.

Recently, an integrated model of enhancers and silencers functions has been proposed (30) which highlights dynamic transitions of promoters between distinct chromatin compartments enriched for functionally-competent enhancer and silencer elements. This model and key elements of the outlined here heterochromatin shuttle hypothesis are beginning to define the time- and chromatin context-dependent 3D view of the interphase chromatin fractal globules dynamics and functions. Evolution of this rapidly emerging field creates highly significant computational and bioinformatics challenges: it would require development and implementation of the next generation systems biology analytical tools and methods to integrate and visualize genome-wide comprehensive profiling data of human transcriptome, DNA sequences and chromatin state maps of individual genomes, annotations of all functional, regulatory, and structural elements within the model of the reference human genome, GWAS data, binding profiles of chromatin fractal globules to nuclear lamina and nucleoli, and extremely large volumes of data from the multiple Hi-C experiments of interphase chromatin folding analyses.

In conclusion, genome-wide analysis of long-range chromatin interactions identifies nucleoli-associated Centromeric Regions of Interphase Chromatin Homing (CENTRICH) on selected human chromosomes which serves as powerful attractors of the interphase chromatin binding, including genetic loci associated with increased risk of developing common human disorders. These findings support the hypothesis that CENTRICH provide scaffolds for 3D structures of the interphase chromosomes by binding chromatin chain clusters, defining important architectural features of the interphase chromatin chain web, and contributing to design of the 3D architecture of human genome. Assortments of proteins bound to CENTRICH domains appear suitable to facilitate chromatin reprogramming, transcriptional competence, and 3D-alignment of interphase chromatin folding by performing continuous genome-wide scans of the interphase chromatin status.

    The essential function of centromeric DNA sequences is to serve as the assembly sites of the kinetochore structures responsible for chromosome movement and segregation during cell division. Identification of CENTRICH strongly argues that centromeric DNA sequences play important role in human cells beyond the critical functions in chromosome segregation and provides essential knowledge for future experimental work on definition of novel molecular and biological functions of centromeric chromatin serving as scaffolds for a 3D landscape of the interphase chromosomes. Specifically, it will be of interest to determine how these unique to human arrangements of centromeric DNA sequences contribute to distinct phenotypic features and inter-individual diversities of *H.sapiens*.

**Materials and Methods**

**Genome-wide analysis of long-range interphase chromatin interactions**. We set out to estimate the density of interphase chromatin binding to a given segment of chromosomes and quantified the number of interphase chromatin binding events per 1 Mb and 1 Kb consecutive segments of each human chromosome. We reasoned that this approach would highlight the genomic regions with distinct ability to attract the homing of interphase chromatin chains by revealing chromosomal segments with markedly increased frequencies of interactions and statistically higher density of interphase chromatin binding sites. In our computational analyses we utilized only sequencing read pairs which were uniquely mapped to the reference human genome as reported in Lieberman-Aiden et al. (13). All Hi-C data were obtained from NCBI GEO (http://www.ncbi.nlm.nih.gov/geo/; GSE189199) and used in their original alignment summaries to the reference human genome. Microsoft Access database containing 59,309,839 records of individual interactions was designed using published Hi-C data which were derived from eight experiments using two human cell lines (13). All analyzed in this work interphase chromatin interactions and Chip-seq data (see below) were defined based on the sequencing tags which are uniquely map to the reference human genome as reported in the original publications. For intergenic genetic markers and regulatory elements such as disease-linked SNPs, snpRNAs, enhancers, a match is called when of the interacting pair coordinates were found within an IDAGL region defined 10 kb up- or downstream of the corresponding disease-associated SNP position. For protein-coding genes, a match is called when at least one of the interacting pair coordinates was found within a gene boundaries or < 1 kb up- or downstream of the corresponding target gene.

**Analysis of Hi-C Data.** Hi-C data from Dekker and colleagues (13) were obtained from NCBI GEO (http://www.ncbi.nlm.nih.gov/geo/; GSE189199); we used the data in their alignment summaries. We compared the lists of interacting chromatin chain pairs to regulator–target gene pairs defined by the genomic coordinates of the intergenic disease-linked snpRNA/enhancer loci, promoter-enhancer interactions, and protein coding genes. To calculate the estimated numbers of interphase chromatin homing sites in each CENTRICH, we analyzed genome-wide chromatin homing data from two biological replicate experiments performed on diploid GM06990 cells and reported these results in the paper. To exclude the potential data bias based on the use the single restriction enzyme, we confirmed these findings by performing the analysis of all 59,309,839 alignment records of individual interactions derived from eight experiments.

**Sequence alignment analysis of human CENTRICH.** We defined as unique genomic sequences the query sequences that have a single identical counterpart in the reference human genome defined by 100% maximal identity with 100% query coverage ad e-value 0 using the MEGABLAST algorithm. All CENTRICH regions reported in this study satisfy this stringent definition of unique DNA sequences in human reference genome database. We carried out the systematic review of the original alignment and mapping data to determine whether DNA sequences of each CENTRICH listed in the Table 1 were defined by the continuous linear alignments of thousands of experimental sequence reads derived from multiple biological replicate experiments. We thought to test whether the definition of unique genomic sequences is true for DNA sequences within the CENTRICH regions by dividing each CENTRICH into series of short continuous segments along the entire length of the region and covering the CENTRICH sequences essentially without any unresolved sequence gaps (Table 2). We utilized the MEGABLAST algorithm to perform the sequences alignment analyses for randomly generated short DNA segments of the CENTRICH regions. All positive experimental regions of mapped sequence tags were grouped by regions of overlap and proximity (within ~ 100 bp), converted into custom track files and examined in the hg18 UCSC genome browser in order to identify overlapped or closely adjacent experimental regions that together represented a single uniquely mapped genomic locus in the reference human genome (Table 2). Coordinates were converted to hg18 and hg19 using the UCSC 'liftOver' utility (http://genome.ucsc.edu/cgi-bin/hgLiftOver).

**Analysis of lamin B1 binding data.** Genome-wide Lamin B1 binding data in human cells were originally reported in Guelen et al. (11) and are available from the original publication and the UCSC genome browser at http://genome.ucsc.edu/ .

**Analysis of nucleoli binding data.** Genome-wide nucleoli binding data in human cells were originally reported in van Koningsbruggen et al. (9) and were obtained from Dr. Lamond (angus@lifesci.dundee.ac.uk).

**Quantitative analyses and visualization of frequencies of interphase chromatin interactions and densities of binding events.** To quantify and visualize these multitudes of binding events genome-wide which govern the folding patterns and 3D relationships of chromatin fractals within the micron size nucleus, we correlated Hi-C method data (eigenvector values) for each chromosome with lamin B1 and nucleoli binding data. Comparison to the Hi-C model was made using 100-kb window eigenvector data corresponding to the checkerboard patterns in Lieberman-Aiden et al. (13). The correlation analyses of the 5 Mb moving averages data sets for each comparison are reported.

    We performed a systematic genome-wide analysis of the distribution of frequencies of long-range chromatin interactions and binding events densities along the lengths of all human chromosomes utilizing two approaches. In the first approach, we computed all uniquely mapped to the reference genome interactions along each chromosome consecutively segregated into 1Mb windows, plotted results of the analysis and look for regions that manifest statistically higher densities

of long-range interactions. For all chromosomal regions displaying statistically significant enrichments of interphase chromatin binding sites within 1 Mb windows, we defined the precise genomic coordinates of the enrichment regions by performing the follow-up analyses at the 100 Kb and 1 Kb resolution windows. In the second approach, we placed all experimentally-defined binding events in ascending orders along the length of each human chromosome, computed average distances between the three consecutive binding sites, normalized values to the genomic distance of 1 Kb, measured this metric using three consecutive binding sites rolling windows, and plotted the results to visualize the distribution of densities of binding sites. Coordinates of genomic locations of reported in this paper CENTRICH determined using these two metrics were undistinguishable.

To assess whether candidate regions of interest manifest densities of binding events greater than expected by chance based on a random distribution model, we used the Fisher's exact test to estimate the statistical significance of the differences in the expected and observed densities of long-range chromatin interactions and/or frequencies of binding events. We accomplished this task by comparing the candidate CENTRICH density values to genome-wide density (test 1); individual chromosomal density (test 2); interaction densities of neighboring regions (test 3; neighboring regions were 5 Mb segments on both sides of the candidate CENTRICH); average value of interaction density for all genome's 1Mb consecutive segments (test 4). Candidate CENTRICH that passes all four statistical tests within all three consecutive resolution window analyses were considered further in subsequent detailed structural-functional analyses (see below) and reported in this paper.

BLAST searches were performed using default search parameters (Expect threshold 10; Word size 32; Match/Mismatch Scores 1,-2) and complete genome sequences of 12.31.2011 freeze (http://blast.ncbi.nlm.nih.gov/Blast.cgi).

**ChIA-PET and RNAPII binding data sets.** Genome-wide data of Chromatin Interaction Analysis with Paired-End-Tag sequencing (ChIA-PET) and mapping of RNAPII binding sites were obtained from Li et al. (14) and used in their original alignment coordinates.

**RNA-Seq data sets.** RNA-Seq datasets were retrieved from the ENCODE data repository site (http://genome.ucsc.edu/ENCODE/) for visualization and analysis of cell type-specific transcriptional activity of defined genomic regions.

**ChIP-Seq and Chip-chip data sets.** The ChIP-Seq and Chip-chip data sets for visualization and analysis were obtained from Rhee and Pugh (15); Robertson et al. (16); Joseph et al. (17), Raha et al. (18), and the ENCODE data repository site (http://genome.ucsc.edu/ENCODE/). We estimated the density of CTCF and STAT1 binding to a given segment of chromosomes by quantifying the number of protein-specific binding events per 1 Mb and 1 Kb consecutive segments of selected human chromosomes and plotting the resulting binding sites density distributions for visualization. To highlight the common and distinct genomic regions with increased propensity to attract the CTCF and STAT1 binding, we correlated the binding sites density distributions of STAT1 and CTCF binding sites. Genome-wide data sets of 35,161 CTCF-bound locations and 52,586 STAT1 binding sites uniquely mapped to the reference human genome were obtained from previously published work (15, 16). We analyzed separately density distributions of 11,004 and 41,582 STAT1 binding sites identified in control and interferon-treated human cells, respectively (16). STAT1 binding genomic locations were identified using ChIP-seq-derived 15.1 and 12.9 million uniquely mapped sequence reads at estimated false discovery rate of less than 0.001 (16).

**Quantitative analysis of chromosome conformation capture assays (3C-qPCR).** The quantitative results of the 3C-qPCR protocol of validation analyses of long-range chromatin complexes from two to four independent preparations of 3C sample with duplicate qPCR data were obtained from Li et al. (14).

**Histone modification and transcription factor ChIP-Seq data sets.** Histone modification and transcription factors ChIP-Seq Data were obtained from the ENCODE data repository site (http://genome.ucsc.edu/ENCODE/) and Li et al. (14). Publicly available histone modification ChIP-Seq libraries from the Broad Institute, the ENCODE Project, and from Joseph et al. (17) targeting the multiple histone modifications (H3K4me1, H3K4me2, H3K4me3, H3K9ac, H3K9me1, H3K20me1, H3K27ac, H3K27me3, and H3K36me3), and transcription factor ChIP-Seq libraries from Raha et al. (18).

**Statistical analyses of the publicly available data sets.** All statistical analyses of the publicly available genomic data sets, including error rate estimates, background and technical noise measurements and filtering, feature peak calling, feature selection, assignments of genomic coordinates to the corresponding builds of the reference human genome, and data visualization were performed exactly as reported in the original publications, described in the ENCODE data web portal and associated references linked to the corresponding data visualization tracks (http://genome.ucsc.edu/ENCODE/). Any modifications or new elements of statistical analyses are described in the corresponding sections of the paper.

**Acknowledgements.**

This work was made possible by the open public access policies of major grant funding agencies and international genomic databases and willingness to share the primary research data by many investigators worldwide. I would like thank you Dr. A. Lamond for providing genome-wide nucleoli binding data in human cells which were originally reported in van Koningsbruggen et al. (9).

**Author Contributions:**

This is a single author contribution. All elements of this work, including conception of ideas, formulation and development of concepts, execution of experiments, analysis of data, and writing the paper, were performed by the author.

**Conflict of Interest:**

No conflict of interest

**Supplement.**

Supplementary Data are available online: Supplementary Text; Supplementary tables S1-S3; Supplementary Figure Legends; Supplementary figures S1-S6.

**Figure legends**

**Figure 1.** Correlation analysis of the Hi-C eigenvector profiles, Lamin B1 binding patterns, and nucleolus attachment sites along the interphase chromatin chains of human chromosomes.

- A. Alignments of the chromosome 2 profiles. Top panel: Inverse correlation of the Hi-C eigenvector profile and Lamin B1 binding pattern. Bottom left panel: Hi-C map of long-range interphase chromatin interactions at 1 Mb resolution. Bottom right panel: alignment of the CENTRICH region to the nucleolus attachment site.

- B. Alignments of the chromosome 17 profiles. Top panel: Inverse correlation of the Hi-C eigenvector profile and Lamin B1 binding pattern. Bottom left panel: Hi-C map of long-range interphase chromatin interactions at 1 Mb resolution. Bottom right panel: alignment of the CENTRICH region to the nucleolus attachment site.

- C. Alignments of the chromosome 10 profiles. Top panel: Inverse correlation of the Hi-C eigenvector profile and Lamin B1 binding pattern. Bottom left panel: Hi-C map of long-range interphase chromatin interactions at 1 Mb resolution. Bottom right panel: alignment of the CENTRICH region to the nucleolus attachment site.

- D. Alignments of the chromosome 1 profiles. Top panel: Inverse correlation of the Hi-C eigenvector profile and Lamin B1 binding pattern. Bottom left panel: Hi-C map of long-range interphase chromatin interactions at 1 Mb resolution. Bottom right panel: alignment of the CENTRICH region to the nucleolus attachment site.

- E. Direct correlation of the strength of association with nucleoli of the centromeric regions of interphase chromatin homing (defined by the values of the nucleoli/genomic ratio indices measured for each chromosome) and the propensity to engage in the long-range inter-chromosomal interactions (defined by the ratios of the inter- to intra-chromosomal interactions measured for each chromosome).

**Figure 2.** Summary of interphase chromatin-binding features of CENTRICH, which function as main attractors of long-range inter-chromosomal homing sites of intergenic disease-associated genomic loci (IDAGL).

A. Graphical illustration of a statistical definition of the genomic coordinates of the interphase chromatin homing sites in defined intergenic regions of human genome on chromosome 2 and chromosome 10 based on a genome-wide Hi-C catalogue of 59,309,839 long-range chromatin interactions reported in Lieberman-Aiden et al. (5). NCBI GEO (http://www.ncbi.nlm.nih.gov/geo/; GSE189199).

B, C. Intergenic disease-associated genomic loci (IDAGL) manifest the overlapping pattern of binding to both nucleolus and nuclear lamina which indicate that these specific genomic regions could alternate (shuttle) between binding to the nucleolus and the nuclear lamina either in different cells within cell population, or at different times within the same cells.

D. Genomic coordinates (human genome build 36.3) of binding sites of the 8q24 gene desert IDAGL within 20 Kb chromosome 2 CENTRICH (top figure), correlation plot (bottom left figure) and linear regression analysis (bottom right figure) showing the inverse correlation between cancer risk odds ratios (OR) defined by the cancer susceptibility risk loci of the 8q24 gene desert and distances between homing sites of the 8q24-locus cancer susceptibility SNPs within chr2 CENTRICH and a middle point of the chr2 CENTRICH. Bars mark genomic coordinates of individual binding sites of corresponding cancer susceptibility SNPs.

E. Genomic coordinates (human genome build 36.3) of the 9p21 gene desert SNPs and intergenic enhancers (13) associated with increased risk of coronary artery disease (CAD) and type 2 diabetes (T2D) and homing sites of the 9p21 locus within 20 Kb chr2 CENTRICH (top figure); correlation plot (bottom left figure) and linear regression analysis (bottom right figure) showing the inverse correlation between CAD and T2D risk odds ratios (OR) defined by the CAD and T2D susceptibility risk loci of the 9p21 gene desert and distances to the nearest homing sites of the 9p21-locus disease susceptibility SNPs within the chr2 CENTRICH. Color-coded bars denote genomic coordinates of disease-associated SNPs, functionally-validated long-range enhancers, and 9p21 locus homing sites to the chr2 CENTRICH.

F. Correlation plot (left figure) and linear regression analysis (right figure) showing the inverse correlation between T2D risk odds ratios (OR) defined by the intronic (6 SNPs) and intergenic (5 SNPs) T2D susceptibility risk loci and distances between their homing sites within chr2 CENTRICH and a middle point of the chr2 CENTRICH. Genomic coordinates of type 2 diabetes risk loci (reviewed in ref. 31) defined by the genome-wide association studies (GWAS) are listed in the Supplemental Table S3.

G. Genomic coordinates (human genome build 36.3) of binding sites of the chromatin chains within the boundaries of the six protein-coding genes listed in Supplemental Table S3 (these genes carry intronic T2D susceptibility SNPs) within 20 Kb chromosome 2 CENTRICH. Blue bars represent individual binding sites. Distinctly colored bars mark gene boundaries and positions of disease-linked SNPs.

**Figure 3.** Topological analysis of the homing sites within chr2 CENTRICH of IDAGLs associated with the increased risk of CAD, T2D, and epithelial malignancies. Analyzed IDAGLs are located in the 9p21 and 8q24 gene deserts, contain multiple disease-linked SNP variants, and harbor functionally-validated long-range enhancers of documented biological relevance to disease phenotypes (see Fig. 2 and text for further details and references).

A. Chromatin state maps of genomic features of the ~ 70 Kb 9p21 gene desert (top) and ~ 1 Mb region (hg 18 coordinates: chr9:21035934-22494089) of validated long-range intergenic enhancers interactions complexes with distant promoters (bottom). Blue ovals denote coordinates of the 26 Kb ECAD7-9 (hg18 coordinates: chr9:22100523-22126469) region which was utilized as the acceptor site in 3C, 3D-DSL, and PCR long-range interactions validation experiments (13). Bracket marks genomic coordinates of the 70 Kb region (hg18 coordinates: chr9:22057440-22127523). Double arrows link validated sites of interactions (blue bars) between the acceptor site within 26 Kb region and donor sites within 1 Mb region. Heights of bars and numbers above bars correspond to the percentage values of the interactions with the donor region as the fraction of all interactions. Single arrows depict genomic position of the CTCF binding sites. Stars denote positions of the STAT1 binding sites.

B. Genomic coordinates of the binding sites of the 70 Kb region of the 9p21 gene desert (see Fig. 2 for details) within chr2 CENTRICH. Arrows connect individual binding sites. Blue oval denotes coordinates of the 26 Kb ECAD7-9 (hg18 coordinates: chr9:22100523-22126469) region which was utilized as the acceptor site in 3C, 3D-DSL, and PCR long-range interactions validation experiments (13).

C. Putative chromatin loop structures deducted from topological analysis of the aligned interacting regions of the 9p21 gene desert and chr2 CENTRICH. Genomic sizes of the chr2 bases and chr9 loops of the putative hetero-chromosomal chromatin chain looping structures defined by the positions of the corresponding binding sites. Percentages denote fractions of all validated long-range enhancers-promoters interactions which occur within particular looping structure.

D. Example of the putative neighboring hetero-chromosomal loops between 9p21 gene desert and chr2 CENTRICH topological structures of which may facilitate 69% of validated long-range enhancers-promoters interactions within 1 Mb chr9 region 22057440-22127523.

E. Topological analysis of the homing sites within chr2 CENTRICH of the 8q24 gene desert (E, F) conferring increased risk of epithelial malignancies (see Fig.2 and text for additional information and references). E: Putative chromatin loop structures deducted from topological analysis of the head to tail alignment of the 8q24 gene desert and chr2 CENTRICH. Genomic sizes of the chr2 bases and chr8 loops of the putative hetero-chromosomal chromatin chain looping structures defined by the positions of the corresponding binding sites. Denoted long-range intergenic enhancers (enh A-G; enh-PC; enh-BC) were functionally validated and many have been shown to physically interact with the *Myc* oncogene. Double arrows link validated sites of interactions between the long-range enhancers and *Myc* oncogene which are topologically compatible with the proposed looping structures. Single arrows depict genomic position of the CTCF binding sites.

F. Putative chromatin loop structures deducted from topological analysis of the head to head alignment of the 8q24 gene desert and chr2 CENTRICH. Genomic sizes of the chr2 bases and chr8 loops of the putative hetero-chromosomal chromatin chain looping structures defined by the positions of the corresponding binding sites. PC, prostate cancer; BC, breast cancer; CRC, colorectal cancer. Denoted long-range intergenic enhancers (enh A-G; enh-PC; enh-BC) were functionally validated and many have been shown to physically interact with the *Myc* oncogene. Double arrows link validated sites of interactions between the long-range enhancers and *Myc* oncogene which are topologically compatible with the proposed looping structures. Single arrows depict genomic position of the CTCF binding sites.

Genomic coordinates of IDAGLs, SNPs, and long-range enhancers were used to identify all homing (binding) sites of the 9p21 and 8q24 gene deserts within the CENTRICH as documented by the Hi-C method. For each gene desert and CENTRICH interaction complex, binding sites density distribution analysis was carried out and results were plotted for visualization, interchromosomal alignments in head-to-head and head-to-tail orientations, and topological compatibility analysis.

**Figure 4.** Topological analysis of the homing sites within chr2 and chr10 CENTRICH of functionally-validated single-gene and multigene interphase chromatin chain clusters with documented long-range enhancer-promoter and/or promoter-promoter interactions and coordinated transcriptional regulation of protein-coding genes. Genomic coordinates of long-range promoter-centered chromatin interactions, which were discovered using ChIA-PET method and confirmed by several independent methods, including qPCR, 3C, and DNA FISH (6), were used to identify all homing (binding) sites of the chromatin chain clusters within the CENTRICH as documented by the Hi-C method. For each chromatin chain cluster and CENTRICH interaction complex, binding sites density distribution analysis was carried out and results were plotted for visualization, interchromosomal alignment in head-to-head and head-to-tail orientations, and topological compatibility analysis. Genomic coordinates and visualization analysis data (shown bottom parts of figures A-C) for long-range enhancers-promoters interactions within each region were obtained from Li et al. (6).

A. Topological analysis of the putative hetero-chromosomal chromatin looping structures, long-range enhancer-promoter interactions, and binding sites of the *IRS1* CAD and T2D risk locus on chr2 within chr10 CENTRICH. Numbers in top figure denote genomic sizes (in Kb) of the chr10 CENTRICH bases (bottom number) and chr2 loops (top numbers) of the putative hetero-chromosomal chromatin chain looping structures which were defined by the positions of the corresponding binding sites. Percentage values denote the number of interactions within the loop region as the fraction of all experimentally-defined interactions. Genomic coordinates of the peaks of the interchromosomal interactions frequency correspond to positions of the high-density binding sites of the *IRS1* locus within the chr10 CENTRICH.

B. Topological analysis of the putative hetero-chromosomal chromatin looping structures, long-range enhancer-promoter interactions, and binding sites of the *SHH* single-gene long-range regulatory enhancer-promoter complex on chr7 within chr2 CENTRICH. Numbers in top figure denote genomic sizes (in Kb) of the chr2 CENTRICH bases (bottom number) and chr7 loops (top numbers) of the putative hetero-chromosomal chromatin chain looping structures which were defined by the positions of the corresponding binding sites. Percentage values denote the number of interactions within the loop region as the fraction of all experimentally-defined interactions. Genomic coordinates of the peaks of the interchromosomal interactions frequency correspond to positions of the high-density binding sites of the *SHH* locus within the chr2 CENTRICH.

C. Topological analysis of the putative hetero-chromosomal chromatin looping structures, long-range enhancer-promoter interactions, and binding sites of the multigene long-range regulatory enhancer-promoter and promoter-promoter complexes on chr14 within chr2 CENTRICH. Numbers in the top part of figure denote genomic sizes (in Kb) of the chr2 CENTRICH bases (bottom number) and chr14 loops (top numbers) of the putative hetero-chromosomal chromatin chain looping structures which were defined by the positions of the corresponding binding sites. Percentage values denote the number of interactions within the loop region as the fraction of all experimentally-defined interactions. Genomic coordinates of the peaks of the interchromosomal interactions frequency correspond to positions of the high-density binding sites of the chr14 multigene regulatory compplexes within the chr2 CENTRICH.

**Figure 5.** CENTRICH represent chromosomal domains containing high density of the CTCF- and STAT1-binding sites.

A. Bar plots of binding sites density distribution analysis of the CTCF (top panels) and STAT1 (bottom panels) on human chr1 (left panels) and chr19 (right panels). Highlighted bars depict values corresponding to the genomic coordinates of CENTRICH.

B. Bar plots of binding sites density distribution analysis of the CTCF (top panels) and STAT1 (bottom panels) on human chr10 (left panels) and chr2 (right panels). Highlighted bars depict values corresponding to the genomic coordinates of CENTRICH.

C. Correlation plots demonstrate significant correlations between genomic locations of the CTCF- and STAT1-binding sites on chr1 (top left figure) and chr19 (top right figure) and density of the STAT1-binding sites within human CENTRICH before and after interferon-gamma treatment (bottom left figure). Linear regression analysis (bottom right figure) documents highly significant correlation between the high density of the CTCF-binding sites within CENTRICH (defined by the Chip-exo method) and increased propensity to attract bining of the interphase chromatin chains (defined by the Hi-C method).

D. Density distribution analysis of the STAT1-binding sites on human chromosome 10 before (top figure) and after (bottom figure) interferon-gamma treatment. Highlighted bars show values within genomic coordinates of the CENTRICH. Inset depicts the correlation plot between the values of the STAT1-binding sites density distribution analysis on human chr10 before and after interferon-gamma treatment.

Genome-wide data sets of the uniquely mapped CTCF- and STAT1-binding sites were obtained from the previously publishes studies (7, 8) and used to carry out binding sites density distribution analysis as described in Methods. Results are shown as the bar plots of the average values of all distances between three consecutive binding sites within chromosomal segments normalized to 1 Kb genomic distance. Bar plots alignments correspond to genomic coordinates of the reference human genome build 36.3.

**Figure 6.** Heterochromatin shuttle model of dynamic features of chromatin fractal globules during the cell cycle.

A. Chromatin chain binding to nuclear lamina (state A) initiates a polymer collapse and folding, promotes spatial segregation of genomic and chromosome territories, and contributes to stability of a fractal globule conformation (folding and compaction cycle). Shuttling between nuclear lamina (state A) and nucleoli (state B) would facilitate chromatin unfolding and loop opening, which would enable a genome-wide cross-talk due to chromatin loop invasion within and across territories and enhance the probability of long-range inter- and intra-chromosomal interactions (unfolding and expansion cycle). A fractal globule conformation would allow a rapid chromatin loop expansion, crossing of boundaries of genomic and chromosome territories, enhancing a likelihood of long-range enhancer/promoter interactions and contributing to formation of transcription factories. Computer-generated images of chromatin fractal globules are adopted from (21).

B. Biomechanical forces created during a two-fold expansion of the nuclear volume in G1, changes of nuclear shape, disassembly of nucleolus, and periodic nuclear rotations would impact the equilibrium of chromatin binding to nuclear lamina and nucleoli and influence initiation, duration, and frequency of a fractal globule loop folding/opening cycles. Computer-generated images of chromatin fractal globules are adopted from (21).

Table 1. Genomic coordinates and interphase chromatin bindings of 4 major human CENTRICH

| CENTRICH | Region START | Region END | Distance, bp | Number of chromatin homing sites per 1 Kb | NCBI Reference Sequence | Fold enrichment of chromatin homing sites | Fisher's exact test two-tail p-values |
|---|---|---|---|---|---|---|---|
| chr2 | 91669025 | 91689823 | 20799 | 1432 | NT_034508.2 | 716 | $1.08E^{-292}$ |
| Chr10_2 | 41916695 | 41921004 | 4309 | 1526 | NT_033985.6 | 763 | $9.47E^{-307}$ |
| chr1 | 121183022 | 121186882 | 3860 | 1189 | NT_077389.3 | 595 | $7.16E^{-255}$ |
| chr10_1 | 41699886 | 41722274 | 22388 | 639 | NT_079540.1 | 320 | $1.36E^{-154}$ |
| chr17 | 22168000 | 22188000 | 20000 | 397 | NT_024862.13 | 199 | $2.10E^{-101}$ |

Legend: Genomic coordinates are from the reference assembly of human genome Homo sapiens (human) (Build 36.3) release. Fisher's Exact Test 2-tail p-values were calculated comparing observed numbers of homing sites to expected number of homing sites based on genome-wide average value of 2 homing sites per 1 Kb.

Table 2. Sequence alignments of the CENTRICH fragments to the human genome reference assembly database using MEGABLAST algorithm.

| CENTRICH segments | Start, NCBI36/hg18 assembly | End, NCBI36/hg18 assembly | Segment size, bp | Number of binding sites within segment | Gap between nearest segments, bp | Reference assembly sequence alignment accession number | Query coverage | E value | Maximum identity |
|---|---|---|---|---|---|---|---|---|---|
| chr2 | 91669000 | 91672892 | 3893 | 3033 | | NT_034508.2 | 100% | 0 | 100% |
| chr2 | 91672915 | 91676328 | 3414 | 3034 | 23 | NT_034508.2 | 100% | 0 | 100% |
| chr2 | 91676425 | 91678077 | 1653 | 2892 | 97 | NT_034508.2 | 100% | 0 | 100% |
| chr2 | 91678123 | 91680582 | 2460 | 1604 | 46 | NT_034508.2 | 100% | 0 | 100% |
| chr2 | 91680607 | 91683282 | 2676 | 4920 | 25 | NT_034508.2 | 100% | 0 | 100% |
| chr2 | 91683308 | 91685690 | 2383 | 4357 | 26 | NT_034508.2 | 100% | 0 | 100% |
| chr2 | 91685709 | 91689823 | 4115 | 7517 | 19 | NT_034508.2 | 100% | 0 | 100% |
| chr17 | 22168150 | 22171017 | 2868 | 861 | | NT_024862.14 | 100% | 0 | 100% |
| chr17 | 22171193 | 22173286 | 2094 | 896 | 176 | NT_024862.14 | 100% | 0 | 100% |
| chr17 | 22173420 | 22176149 | 2730 | 880 | 134 | NT_024862.14 | 100% | 0 | 100% |
| chr17 | 22176189 | 22178092 | 1904 | 1988 | 40 | NT_024862.14 | 100% | 0 | 100% |
| chr17 | 22178106 | 22180328 | 2223 | 807 | 14 | NT_024862.14 | 100% | 0 | 100% |
| chr17 | 22180391 | 22183308 | 2918 | 970 | 63 | NT_024862.14 | 100% | 0 | 100% |
| chr17 | 22183396 | 22187058 | 3664 | 1460 | 88 | NT_024862.14 | 100% | 0 | 100% |
| chr10 | 41703254 | 41703594 | 341 | 134 | | NT_033985.7 | 100% | 2.00E-178 | 100% |
| chr10 | 41703612 | 41704785 | 1174 | 297 | 18 | NT_033985.7 | 100% | 0 | 100% |
| chr10 | 41704901 | 41705268 | 368 | 707 | 116 | NT_033985.7 | 100% | 0 | 100% |
| chr10 | 41705361 | 41706034 | 674 | 477 | 93 | NT_033985.7 | 100% | 0 | 100% |
| chr10 | 41706068 | 41707244 | 1177 | 253 | 34 | NT_033985.7 | 100% | 0 | 100% |
| chr10 | 41707251 | 41707439 | 189 | 149 | 7 | NT_033985.7 | 100% | 4.00E-94 | 100% |
| chr10 | 41707677 | 41708496 | 820 | 424 | 238 | NT_033985.7 | 100% | 0 | 100% |
| chr10 | 41708562 | 41709519 | 958 | 304 | 66 | NT_033985.7 | 100% | 0 | 100% |
| chr10 | 41709521 | 41711880 | 2360 | 480 | 2 | NT_033985.7 | 100% | 0 | 100% |
| chr10 | 41712011 | 41714205 | 2195 | 662 | 131 | NT_033985.7 | 100% | 0 | 100% |
| chr10 | 41714256 | 41714993 | 738 | 196 | 51 | NT_033985.7 | 100% | 0 | 100% |
| chr10 | 41715206 | 41716855 | 1650 | 630 | 213 | NT_033985.7 | 100% | 0 | 100% |
| chr10 | 41716968 | 41719197 | 2230 | 977 | 113 | NT_033985.7 | 100% | 0 | 100% |
| chr10 | 41719208 | 41722274 | 3067 | 5887 | 11 | NT_033985.7 | 100% | 0 | 100% |
| chr10 | 41916812 | 41920515 | 3704 | 2292 | 194538 | NT_033985.7 | 100% | 0 | 100% |
| chr1 | 121186000 | 121186882 | 883 | 3569 | | NT_077389.3 | 100% | 0 | 100% |

**Table 3.** Enrichment of CTCF-binding sites in human CENTRICH.

| CENTRICH chromosome | Number of CTCF-binding sites | CENTRICH size, bp | Fold enrichment | P value |
|---|---|---|---|---|
| chr1 | 7 | 3860 | 185.6733 | 2.53E-12 |
| chr2 | 6 | 20799 | 29.53572 | 2.04E-07 |
| chr4 | 2 | 77063 | 2.657192 | 0.17797 |
| chr7 | 3 | 15513 | 19.79996 | 0.000781 |
| chr10_1 | 13 | 22388 | 59.45205 | 7.87E-18 |
| chr10 spacer | 7 | 194421 | 3.686325 | 0.003692 |
| chr10_2 | 2 | 4309 | 47.52174 | 0.001368 |
| chr12 | 1 | 400271 | 0.255791 | 0.197144 |
| chr16 | 1 | 22686 | 4.513162 | 0.208064 |
| chr19 | 12 | 17065 | 71.99689 | 2.99E-17 |
| Genome-wide | 35,161 | 3.6E+09 | 1 | 1 |

Legend: Numbers of CTCF-binding sites within human CENTRICH were calculated using previously published genome-wide data sets (7). Fisher's exact test 2-tail p values were calculated by comparisons with the genome-wide estimates based on a random distribution hypothesis (http://www.langsrud.com/fisher.htm).

Table 4. Enrichment of STAT1-binding sites in human CENTRICH.

| CENTRICH Chromosome | Number of STAT1 sites, IFN (-) | Number of STAT1 sites, IFN (+) | Fold enrichment IFN (-) | P value | Fold enrichment IFN (+) | P value | Number of sites, % change | Enrichment, % change |
|---|---|---|---|---|---|---|---|---|
| chr2 | 11 | 11 | 173.0035 | 2.58E-20 | 45.78578 | 4.46E-14 | 0 | -73.5348 |
| Chr10_2 | 1 | 2 | 75.91508 | 0.015146 | 40.18222 | 0.001897 | 100 | -47.0695 |
| chr1 | 2 | 1 | 169.4912 | 0.000138 | 22.42813 | 0.055806 | -50 | -86.7674 |
| chr10_1 | 4 | 4 | 58.44525 | 1.22E-06 | 15.46768 | 0.000208 | 0 | -73.5348 |
| chr17 | 0 | 2 | 0 | 1 | 8.657259 | 0.025889 | UP | UP |
| chr7 | 11 | 9 | 231.9538 | 2.64E-21 | 50.22583 | 5.72E-12 | -18.1818 | -78.3466 |
| chr19 | 1 | 1 | 19.16895 | 0.053459 | 5.073108 | 0.186755 | 0 | -73.5348 |
| chr16 | 20 | 16 | 288.3876 | 4.37E-39 | 61.05799 | 2.46E-21 | -20 | -78.8278 |
| chr21 | 4 | 3 | 16.06887 | 0.000144 | 3.1895 | 0.071894 | -25 | -80.1511 |
| chr4 | 21 | 18 | 89.14109 | 1.51E-32 | 20.2212 | 5.26E-17 | -14.2857 | -77.3155 |
| chr12 | 13 | 13 | 10.62414 | 8.58E-10 | 2.811704 | 0.001128 | 0 | -73.5348 |
| All CENTRICH | 88 | 80 | 42.00503 | 3.08E-105 | 10.10541 | 1.14E-50 | -9.09091 | -75.9424 |
| Genome-wide | 11004 | 41582 | 1 | 1.00E+00 | 1 | 1 | 277.8808 | 277.8808 |
| chr10 spacer | 8 | 8 | 13.4602 | 2.64E-07 | 3.562273 | 0.002454 | 0 | -73.5348 |

Legend: Numbers of STAT1-binding sites within human CENTRICH were calculated using previously published genome-wide data sets (8). IFN, interferon-$\gamma$. Fisher's exact test 2-tail p values were calculated by comparisons with the genome-wide estimates based on a random distribution hypothesis (http://www.langsrud.com/fisher.htm).

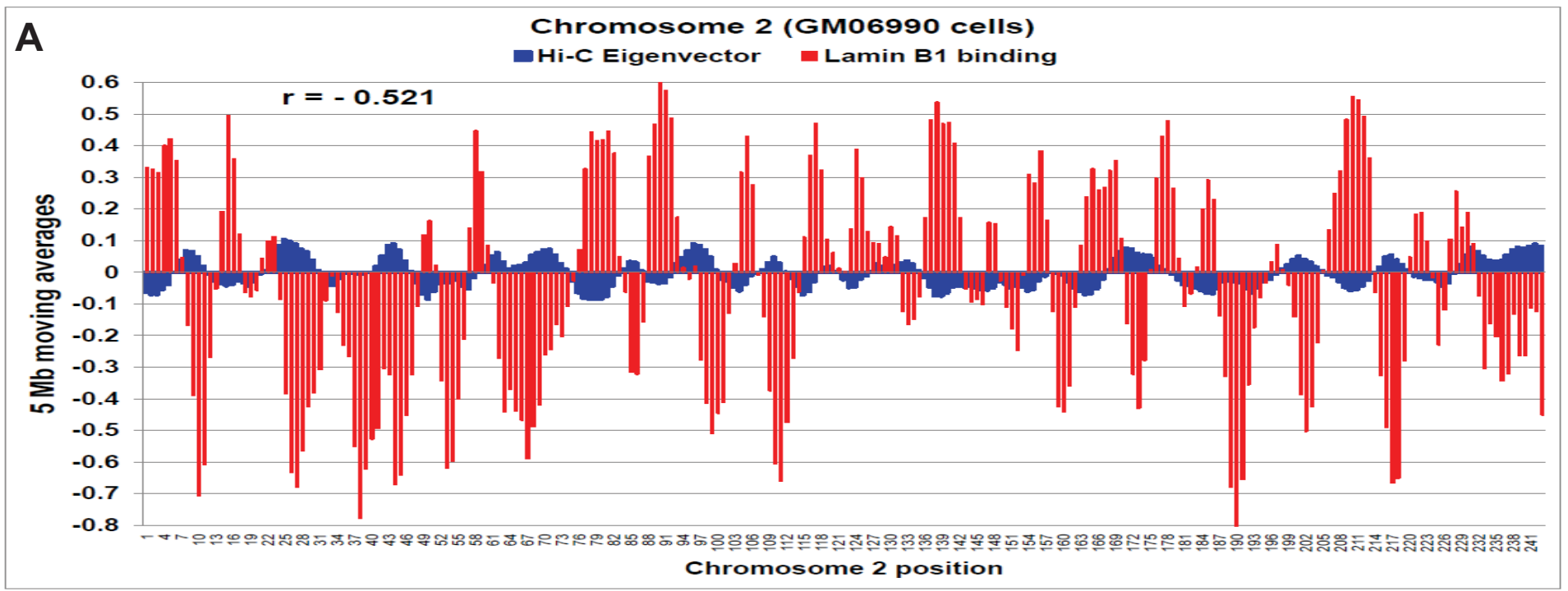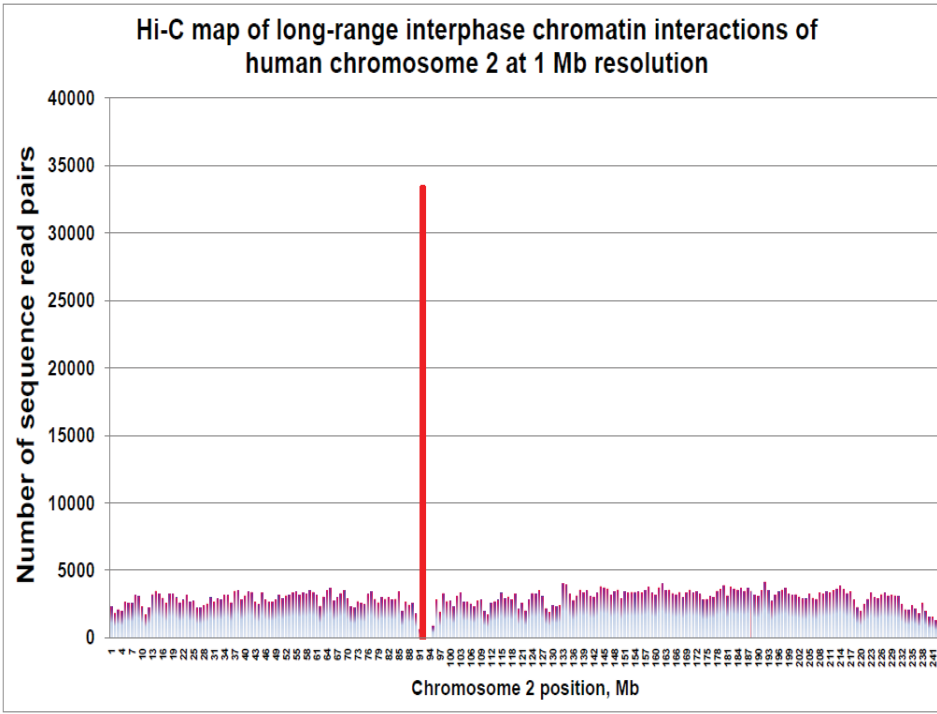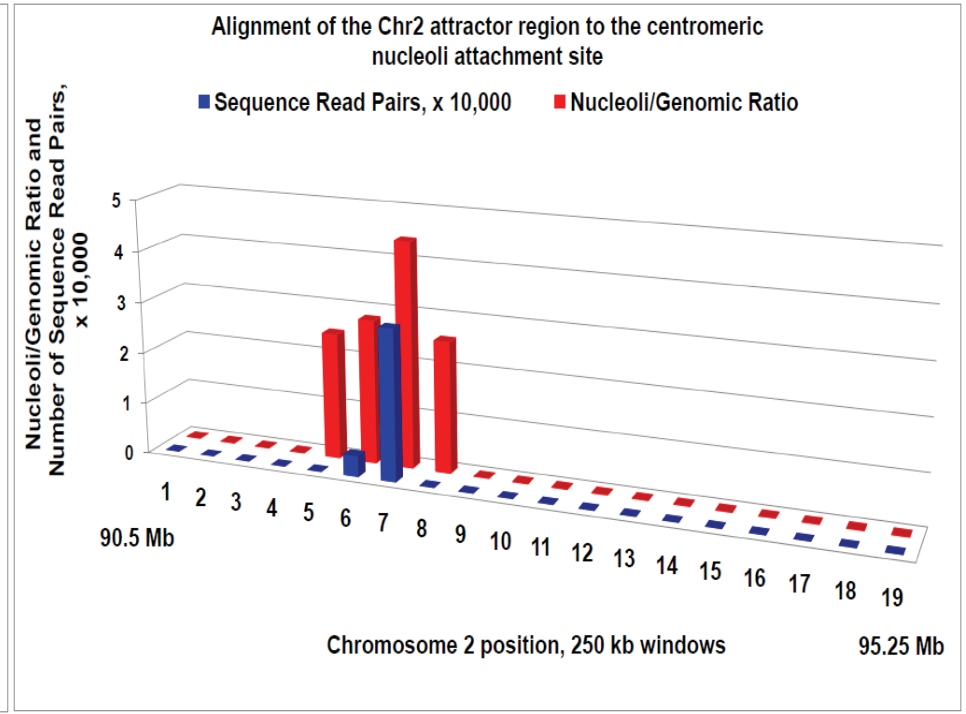

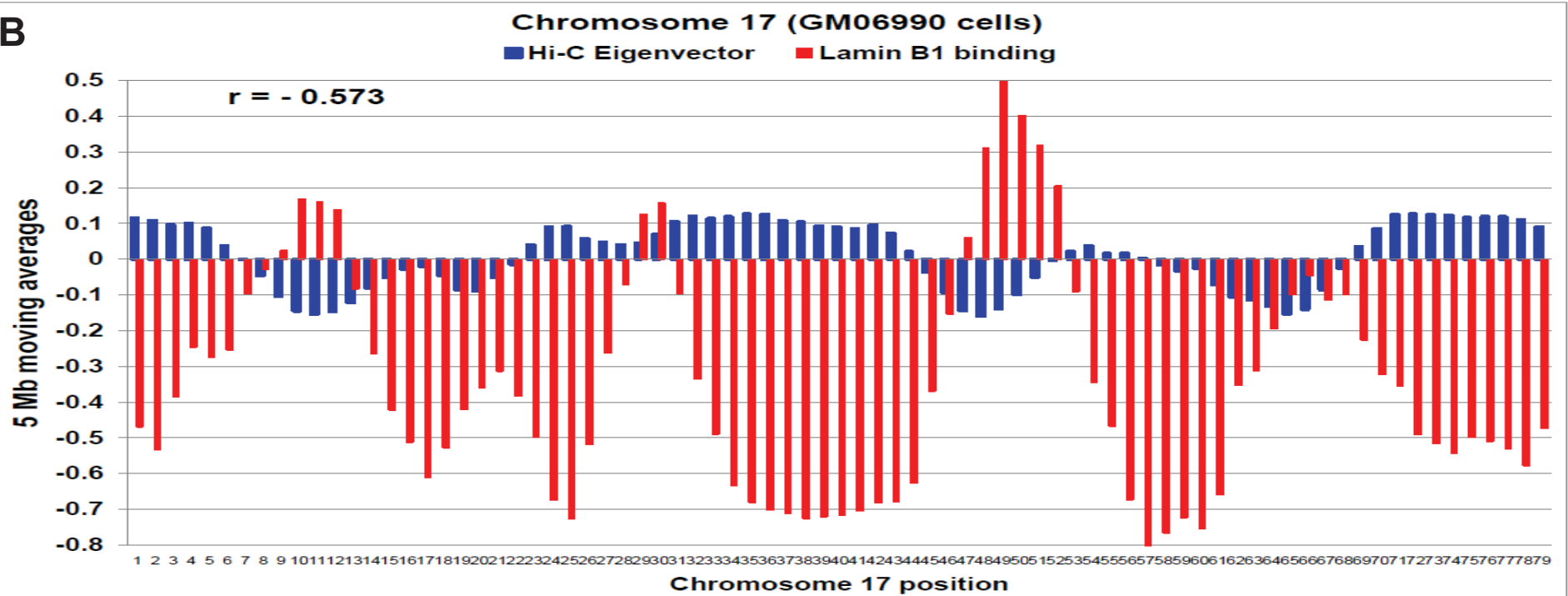
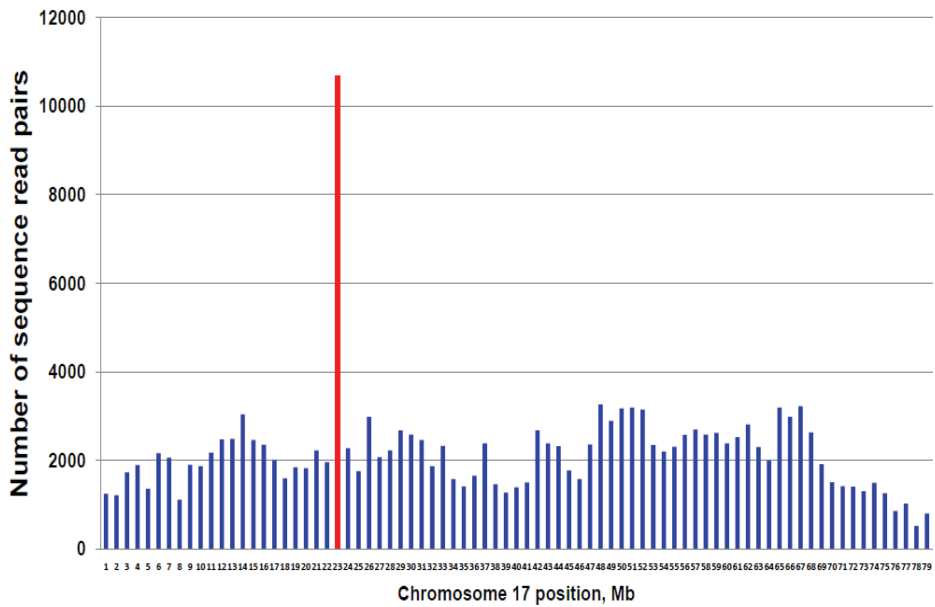
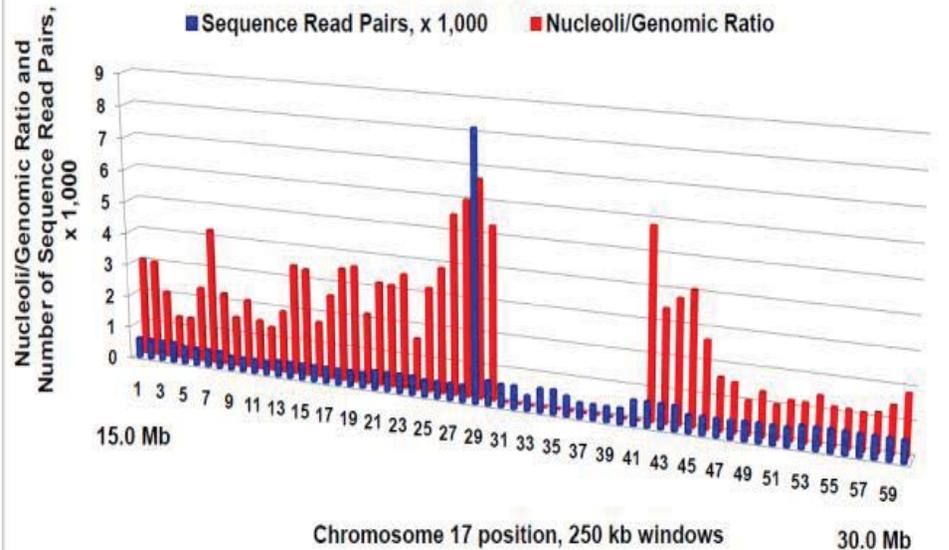

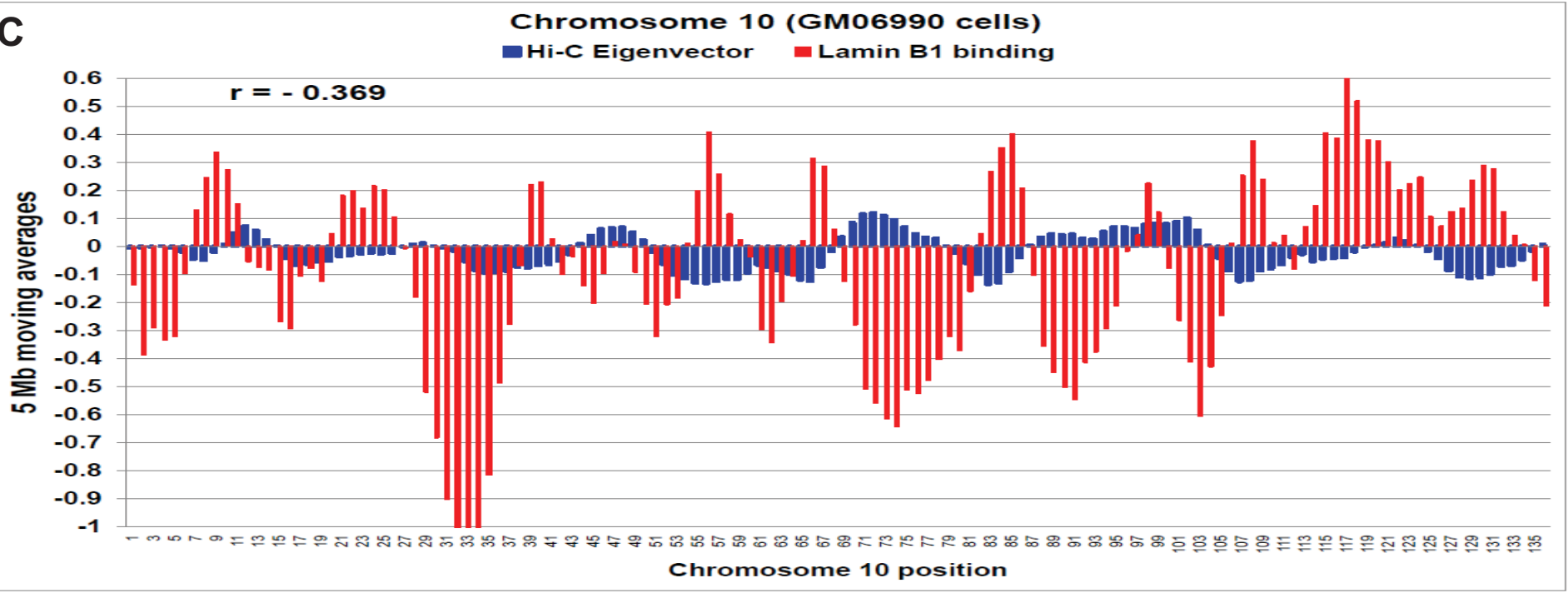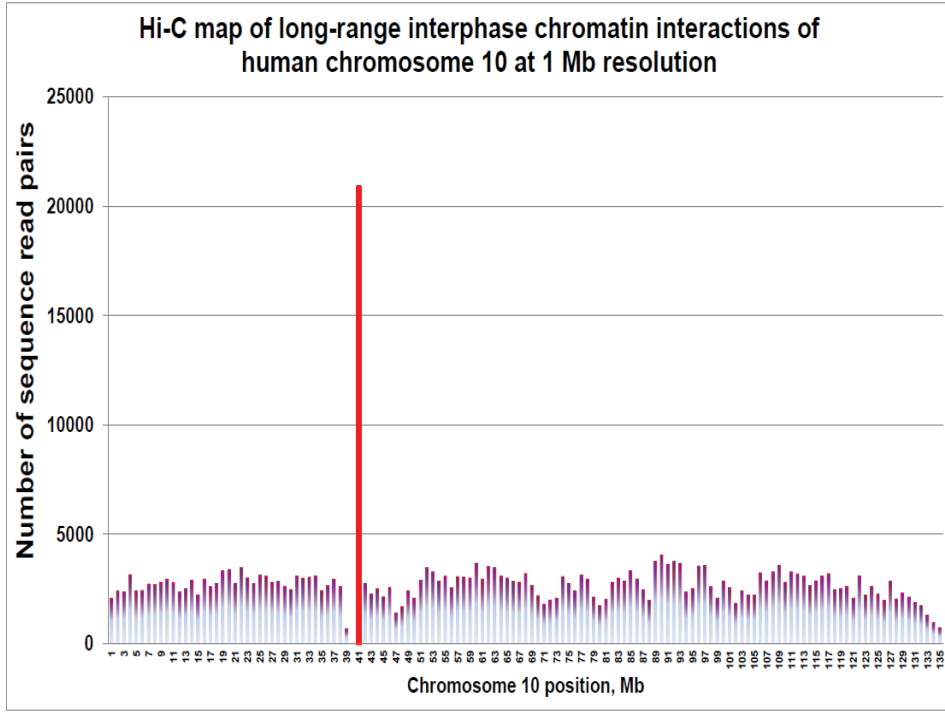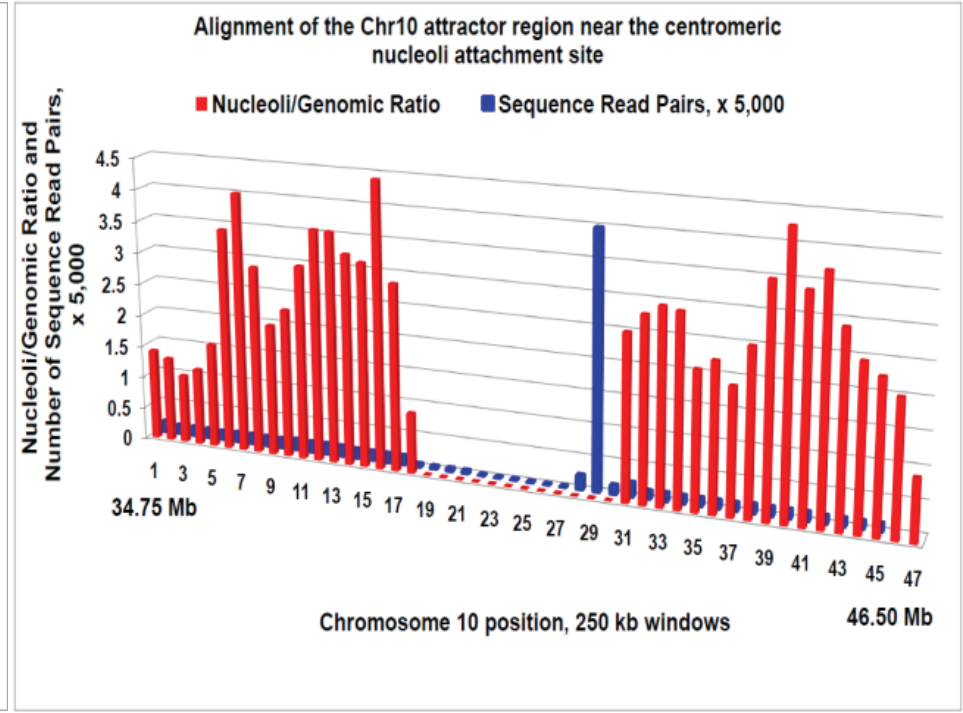

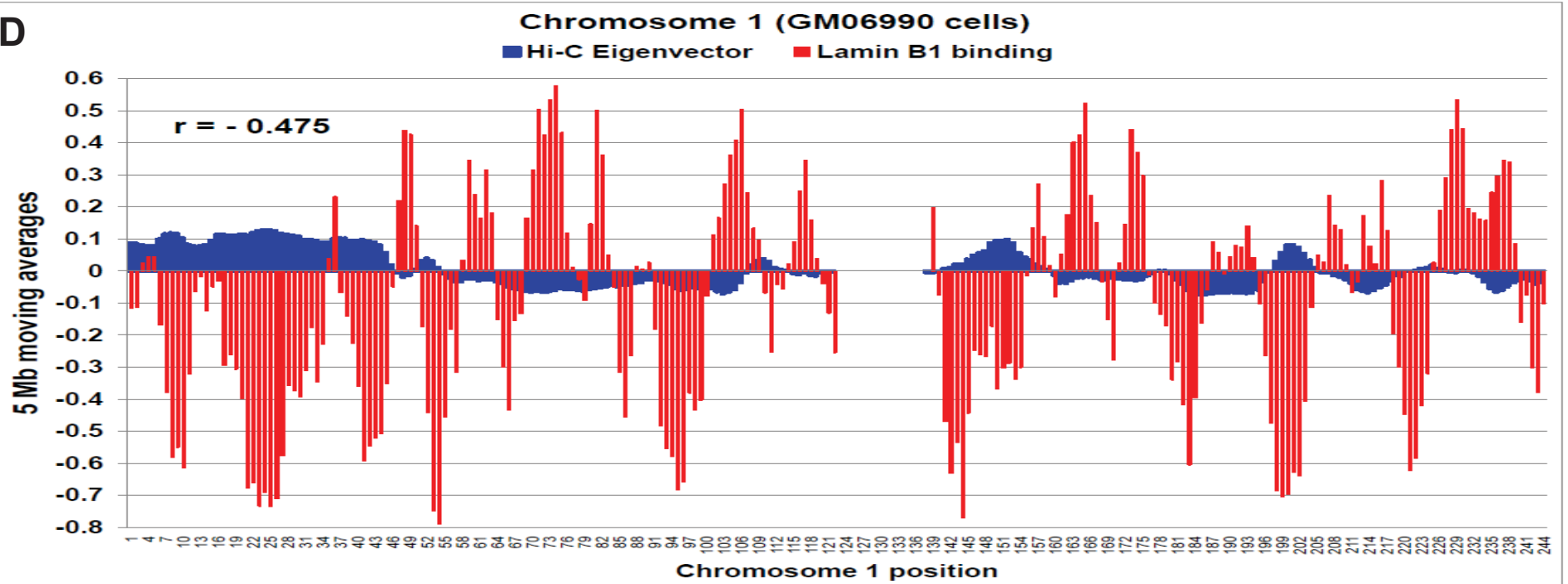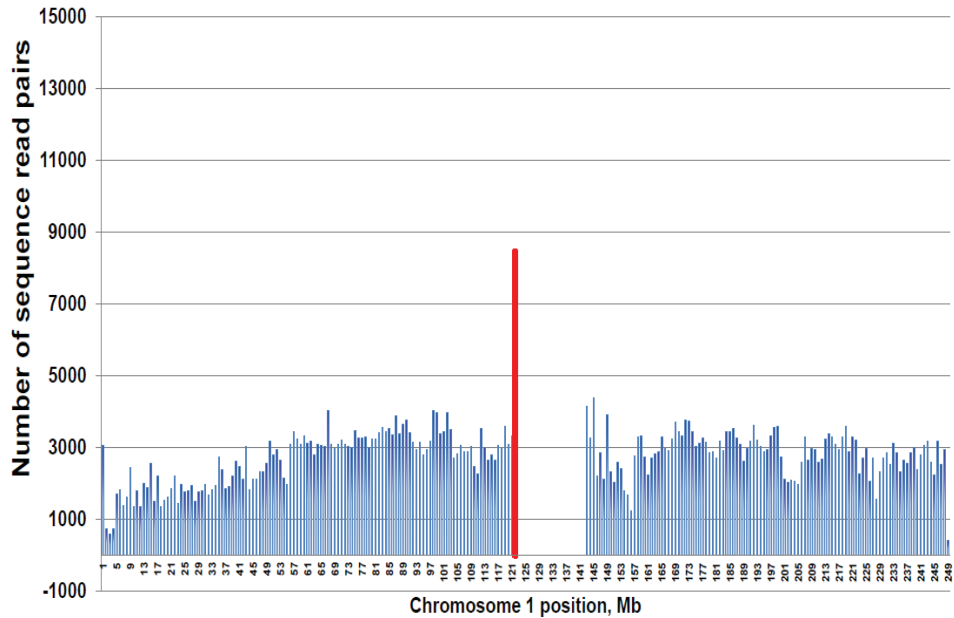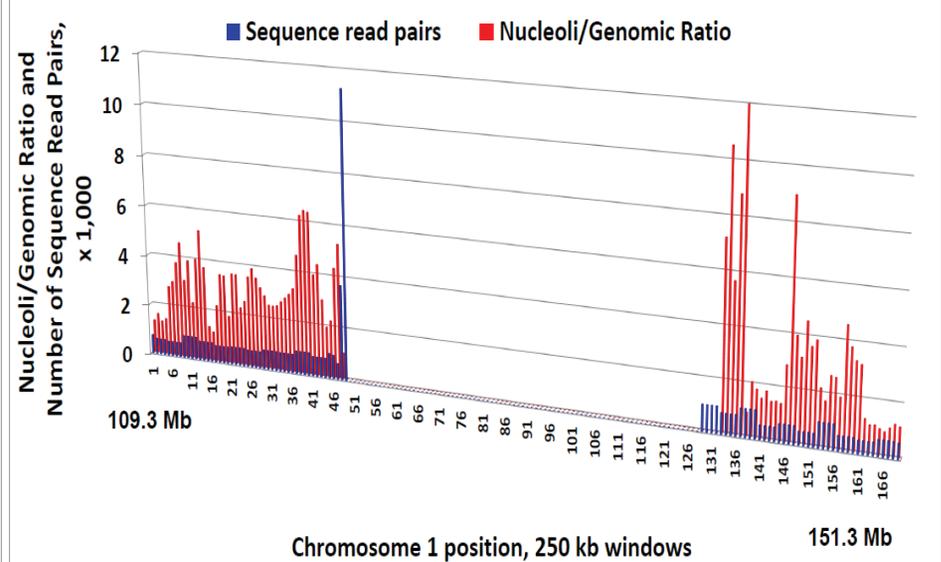

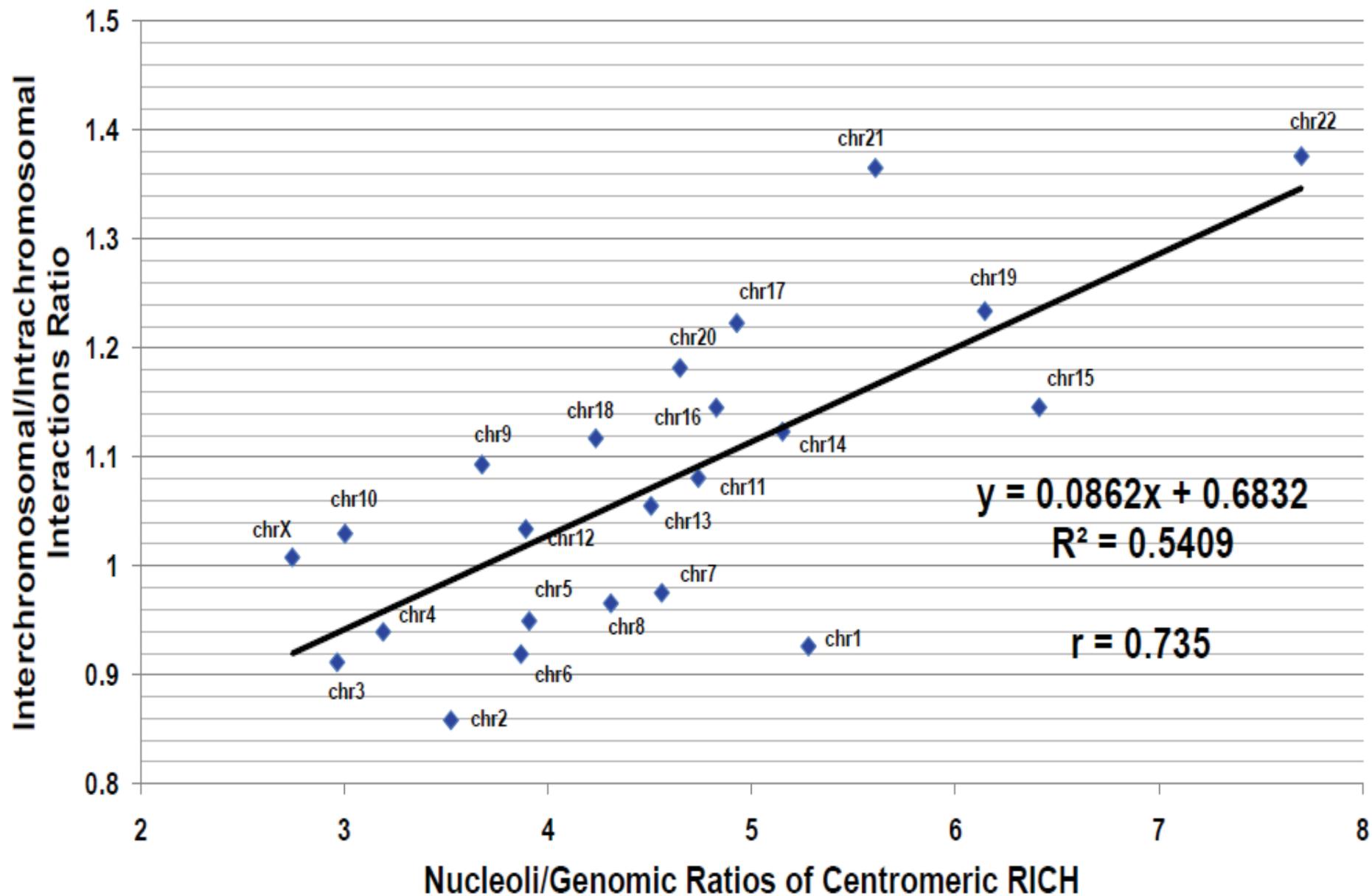

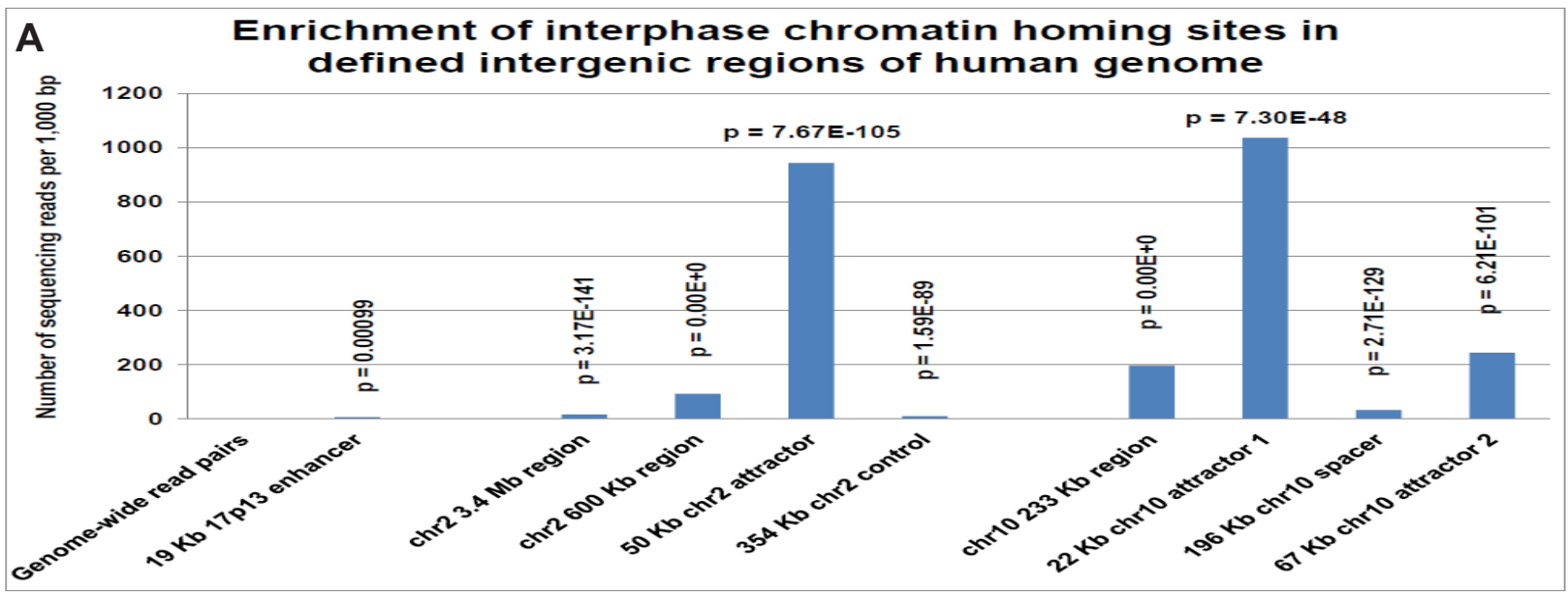

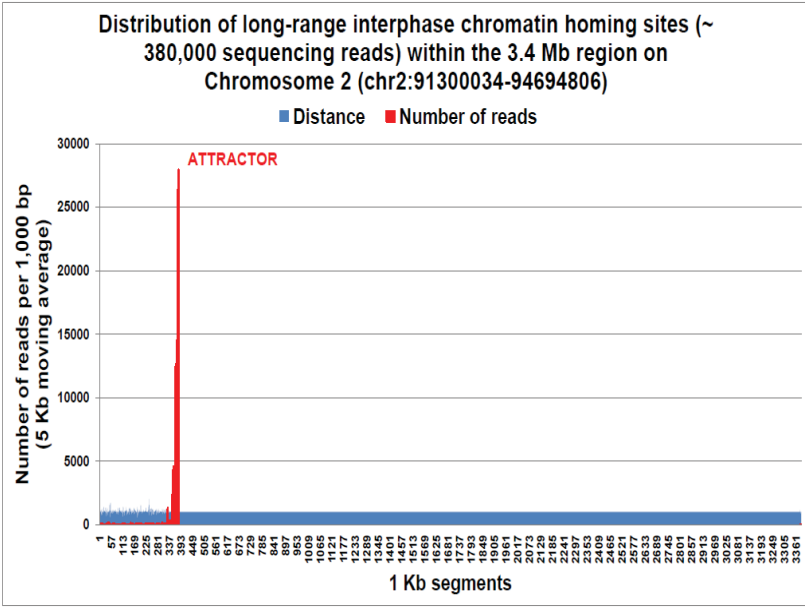
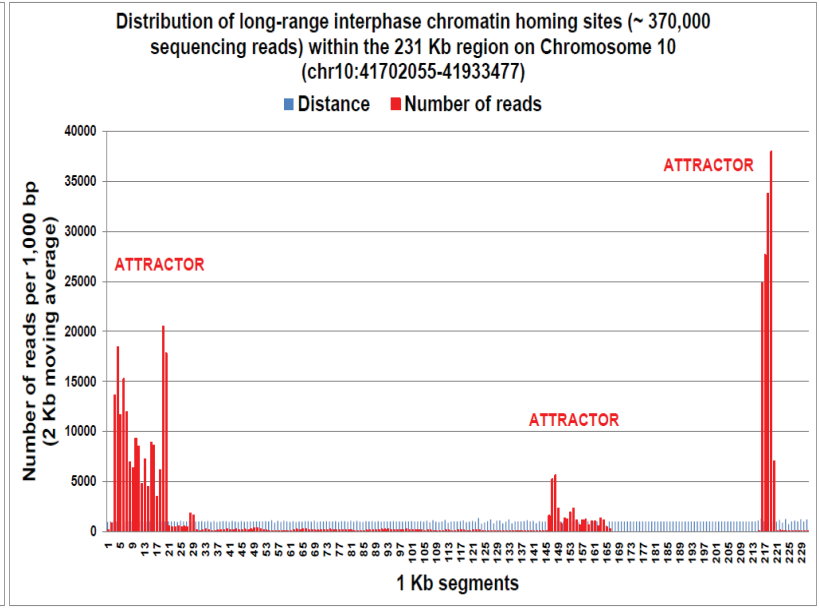

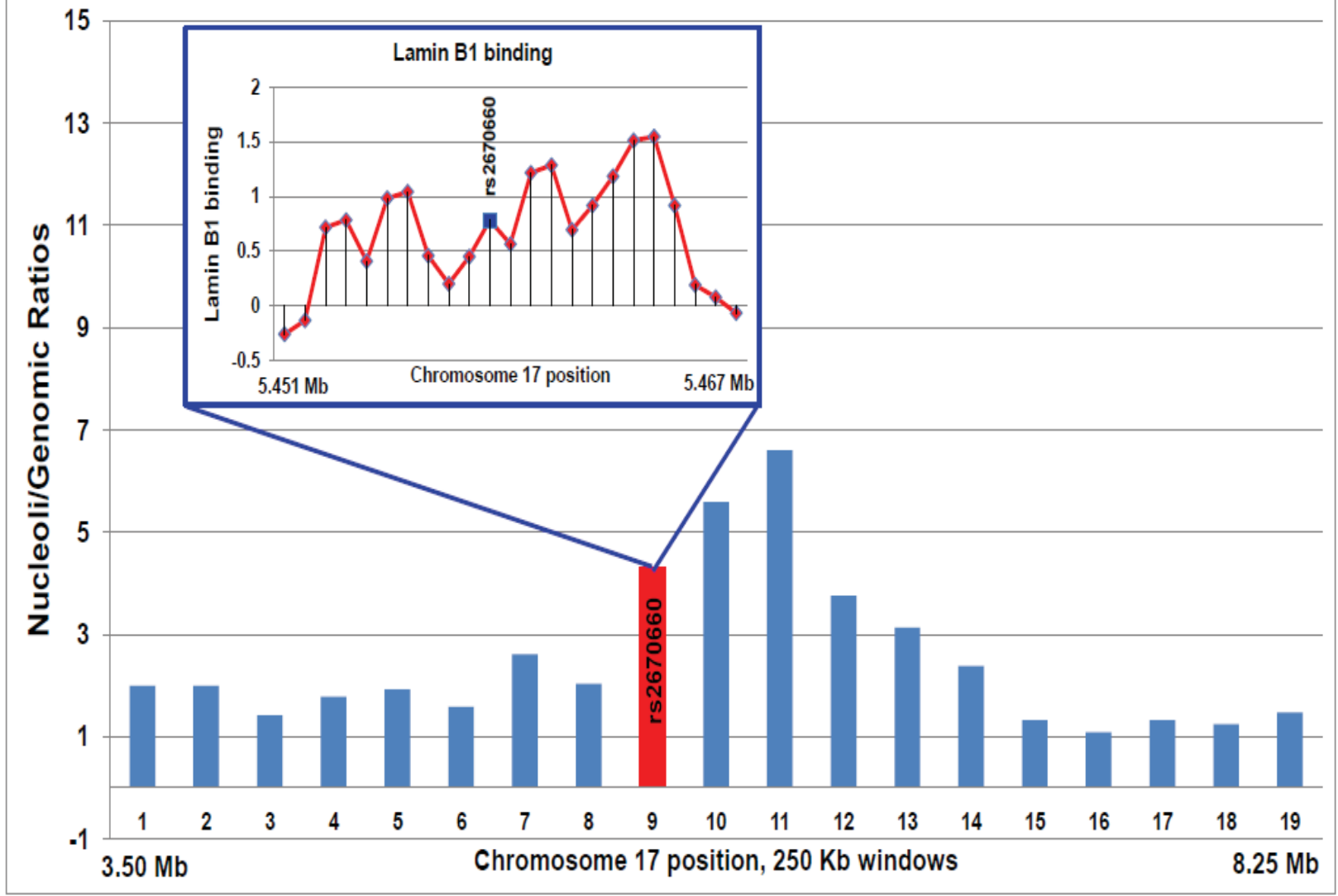

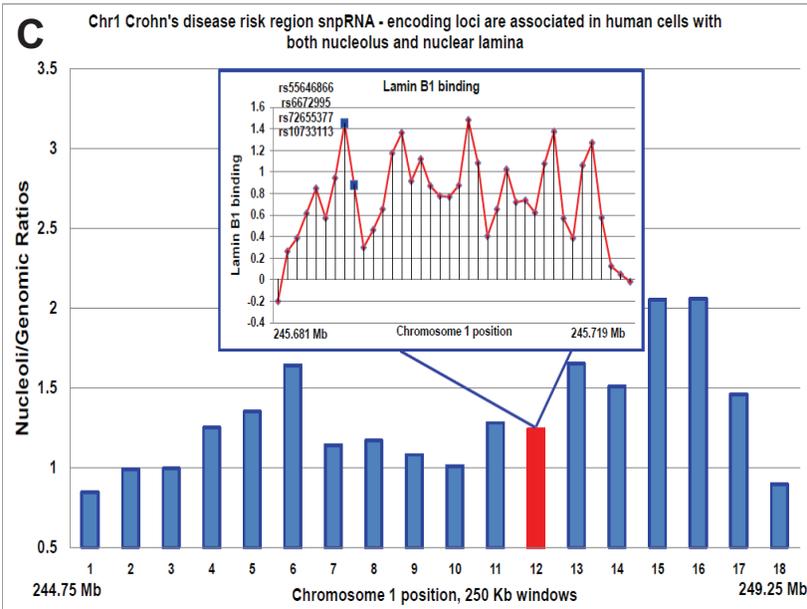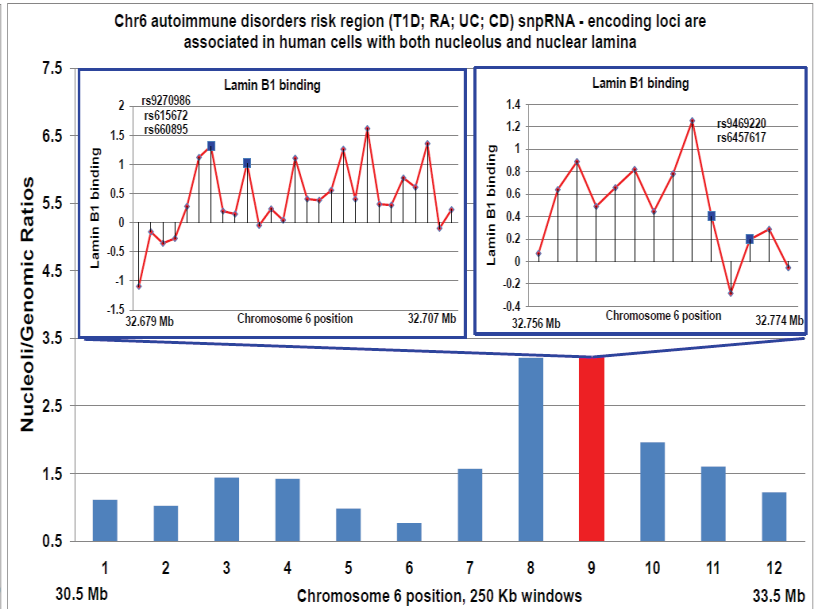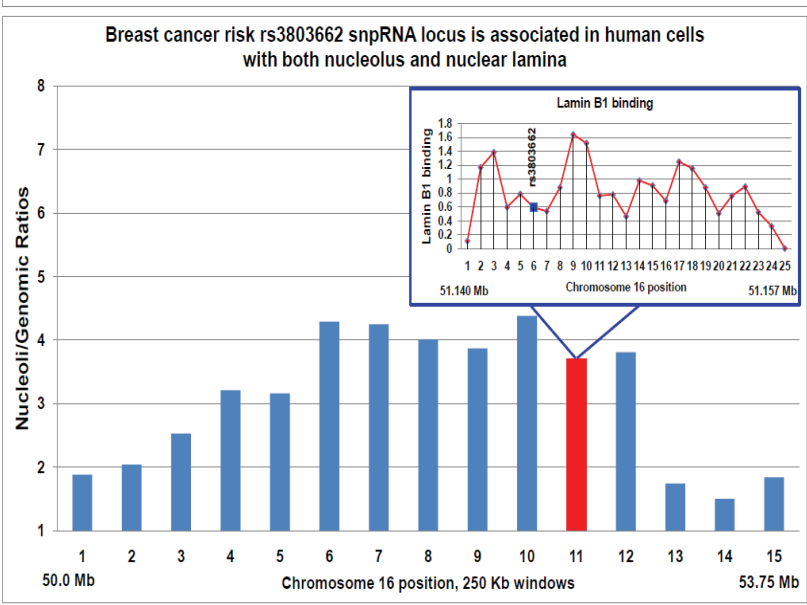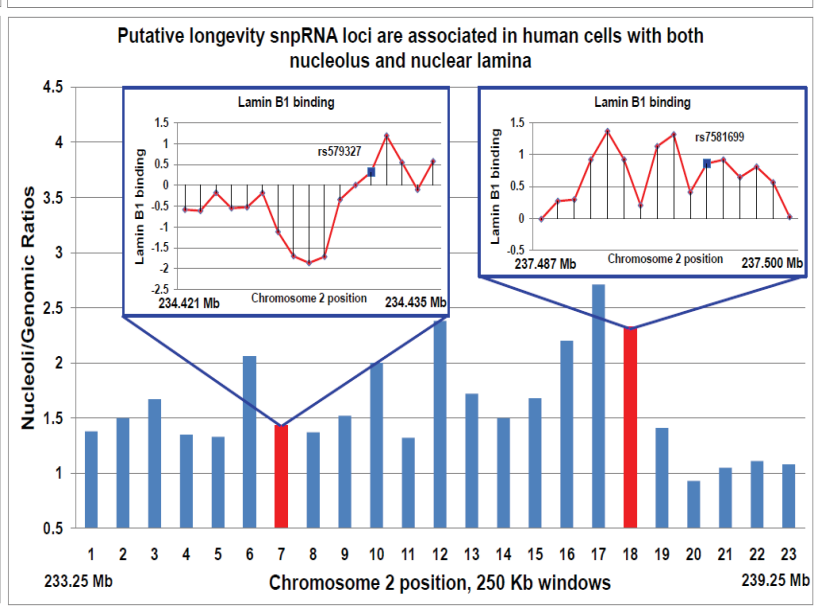

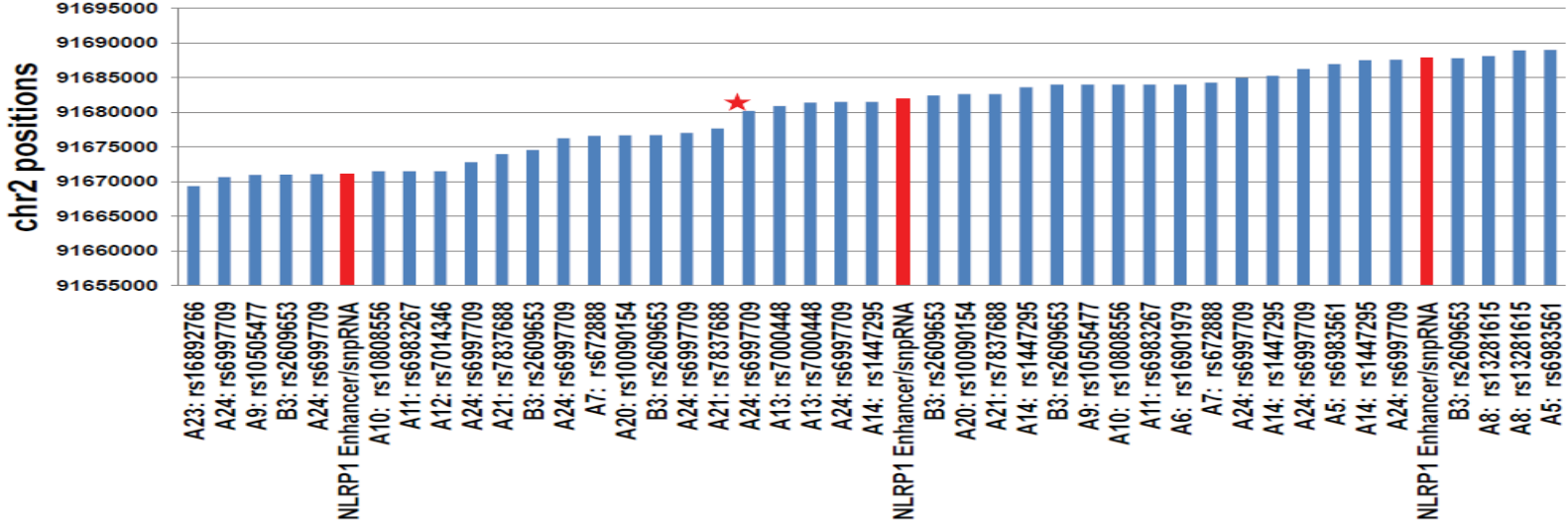
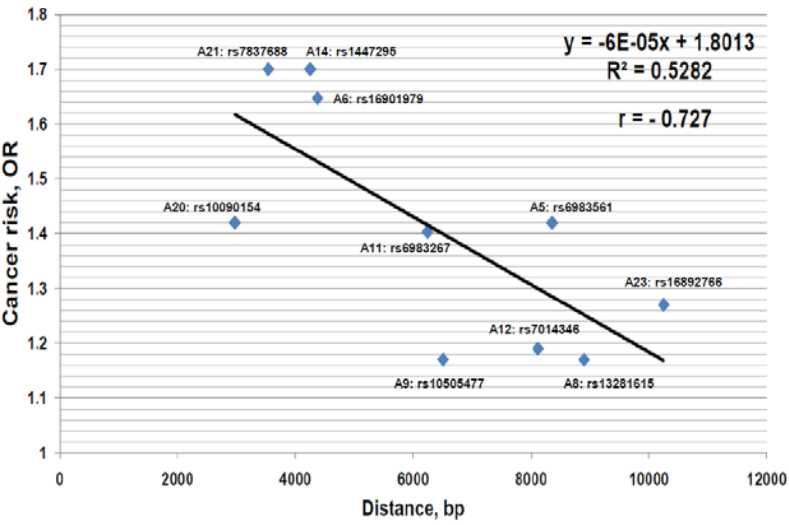
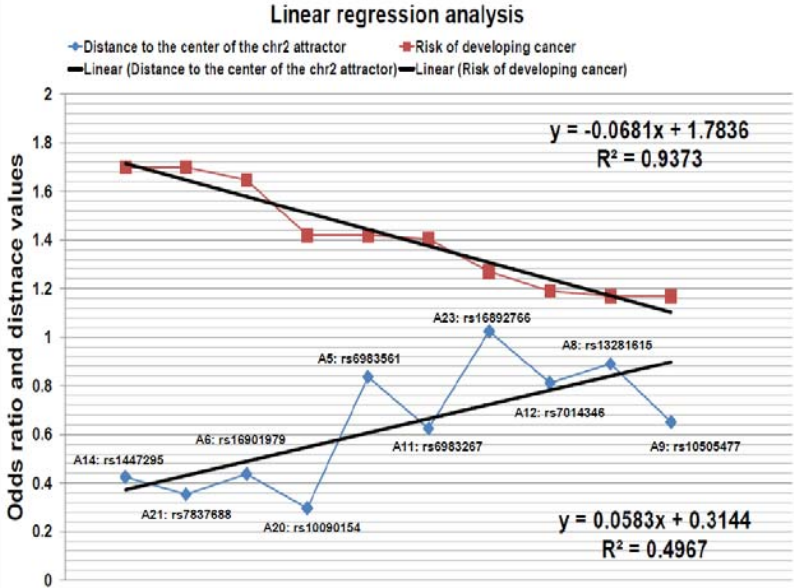

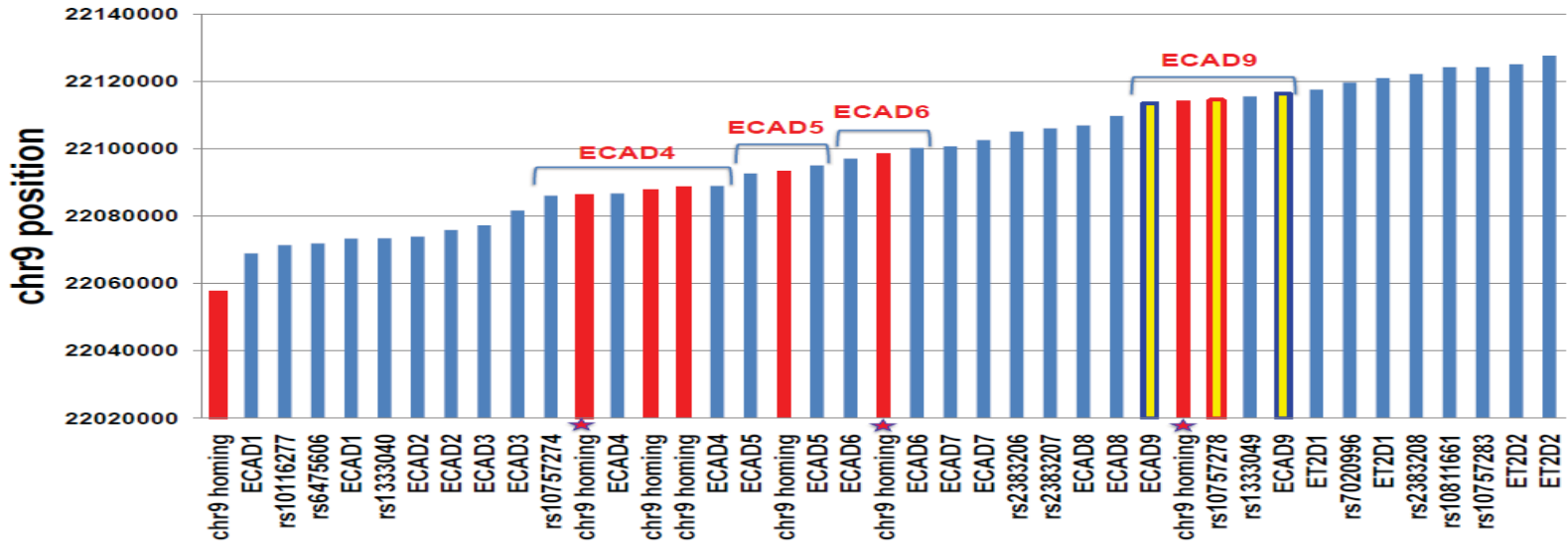
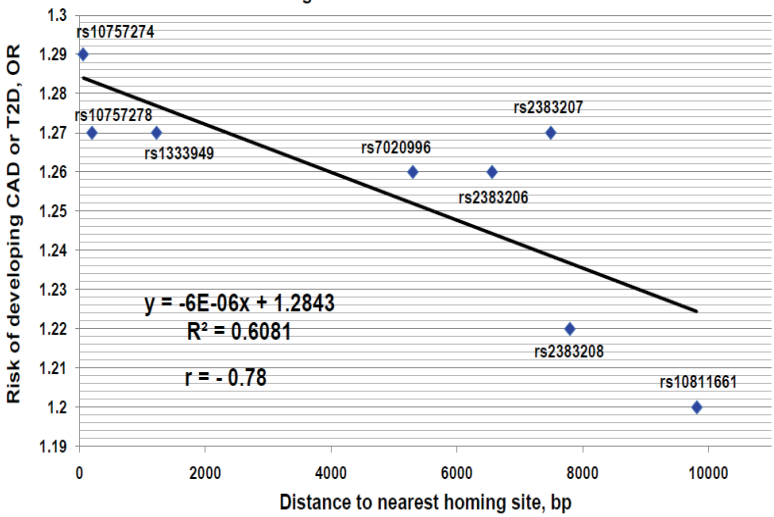
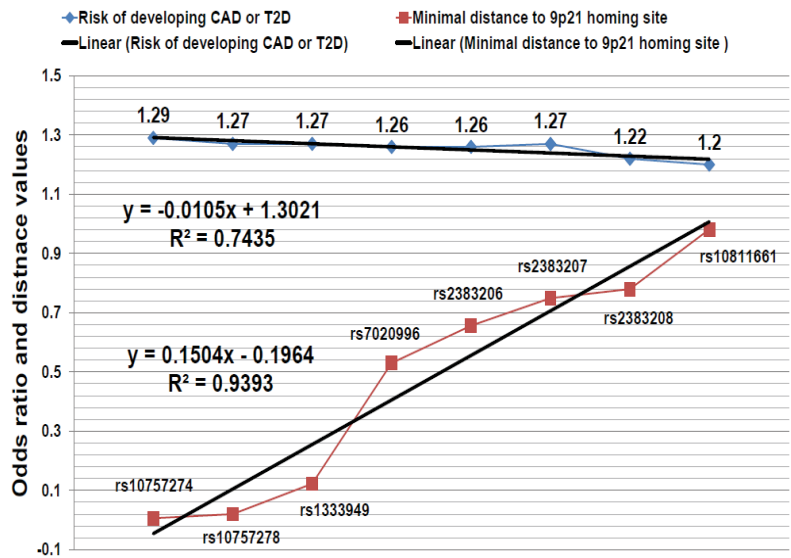

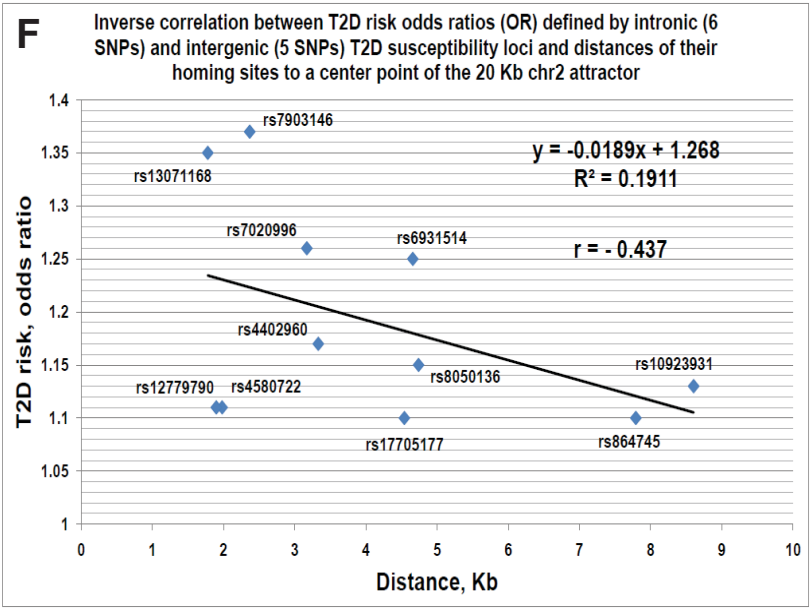
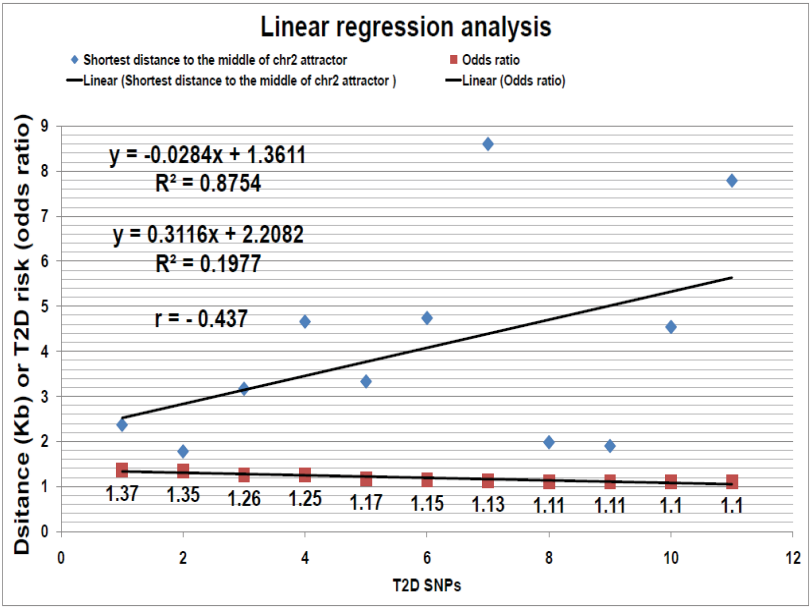
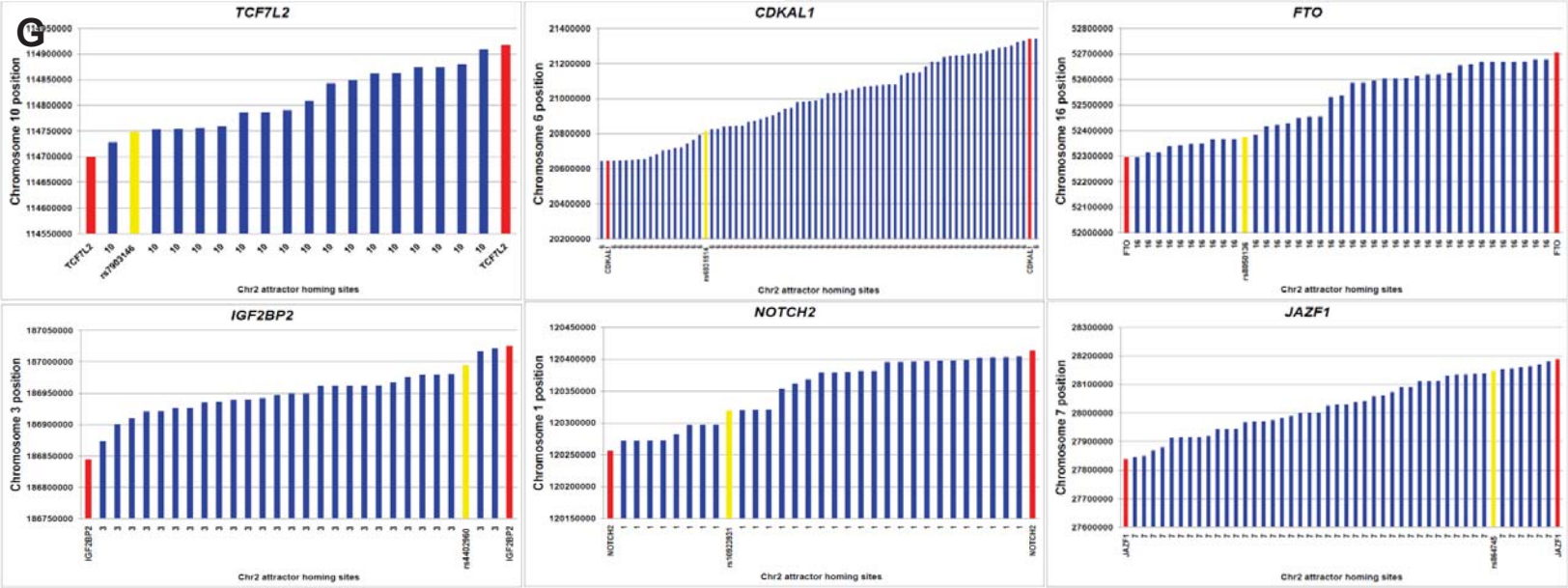

A

chr9:22,057,440-22,127,523 (size 70,084 bp.)

ECAD7-9
chr9:22,100,523-22,126,469
(size 25946 bp.)

IFNA21    MTAP    CDKN2A/B    ECAD6    DOTAR

chr9:21,152,278-22,246,919 (size 1,094,642 bp.)

**B**

Genomic coordinates of the 9p21 gene desert SNPs and enhancers associated with increased risk of CAD and T2D and homing sites within the 20 Kb chr2 attractor

STAT1 binding enrichment

CTCF/STAT1 binding enrichment

Chr2 CENTRICH

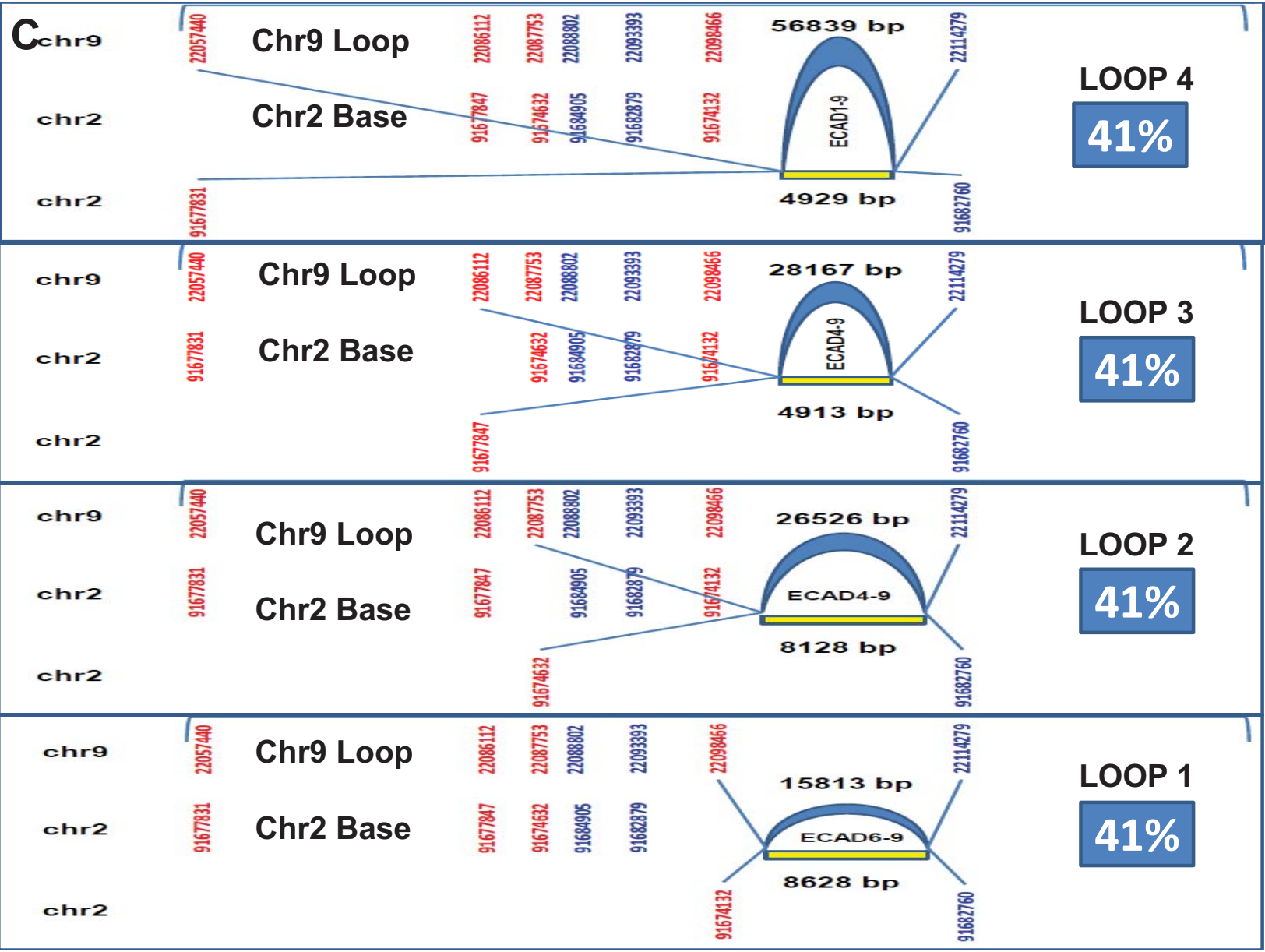

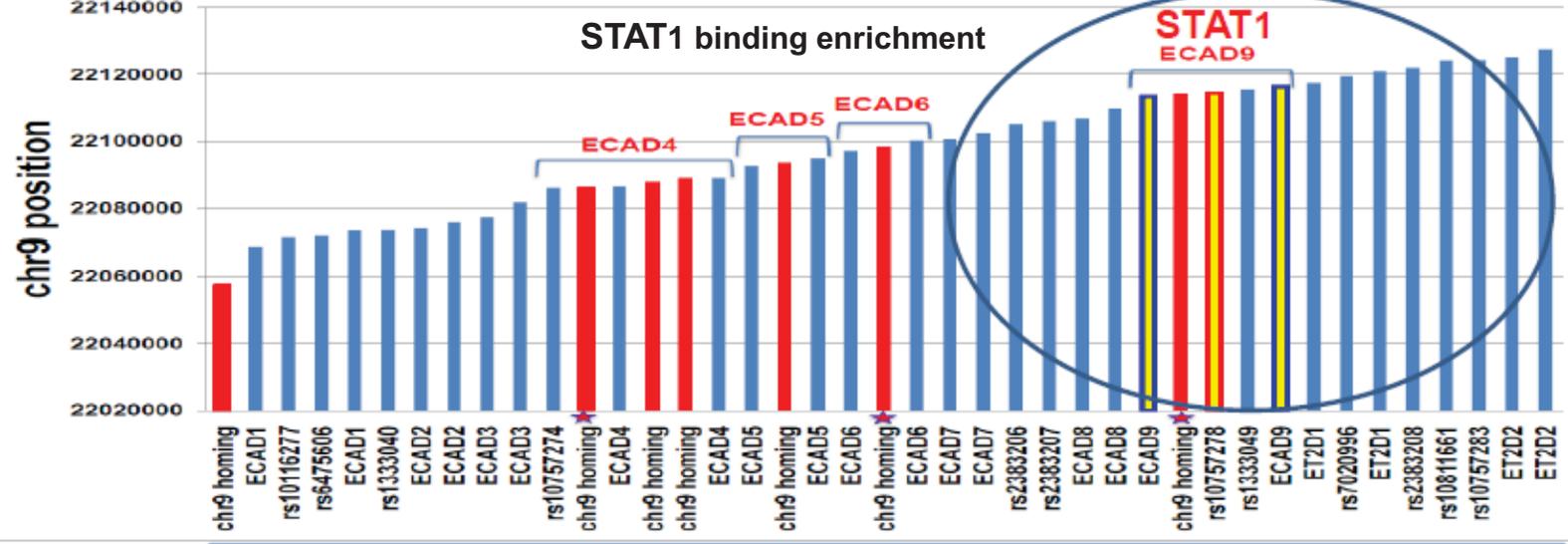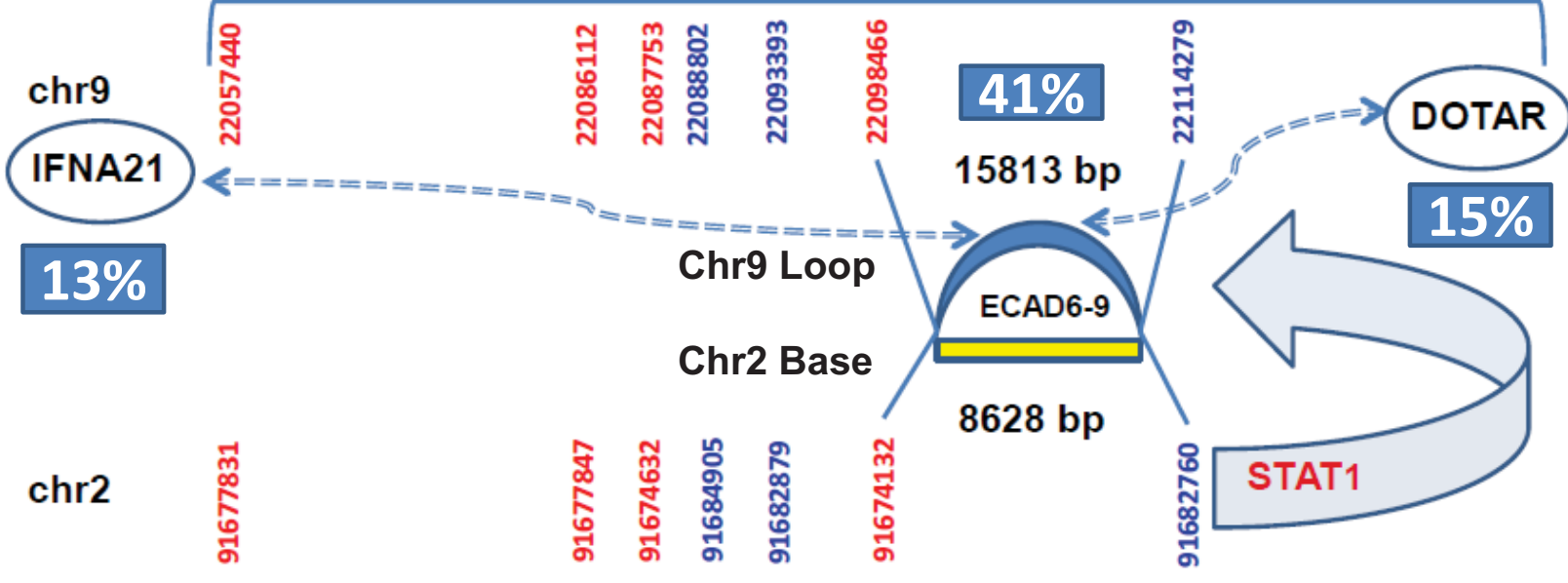

# E

## Intergenic long-range enhancers and cancer risk SNPs within 8q24 gene desert

Prostate cancer (128,174 kb)
Breast cancer (128,425 kb)
Prostate cancer Colorectal cancer (128,482 kb)
Prostate cancer (128,600 kb)

| | chr8 | chr2 |
|---|---|---|
| | 128172275 | 91688975 |
| | 128181448 | 91686968 |
| | 128202385 | 91683988 |
| | 128408376 | 91684270 |
| | 128411683 | 91676610 |
| | 128430800 | 91688114 |
| | 128431621 | 91688914 |
| | 128475779 | 91670964 |
| | 128481956 | 91683982 |
| | 128490880 | 91671493 |
| | 128505246 | 91680884 |
| | 128514508 | 91681396 |
| | 128553537 | 91683607 |
| | 128557010 | 91687521 |
| | 128557645 | 91685280 |
| | 128558232 | 91681515 |
| | 128599270 | 91676681 |
| | 128608075 | 91682628 |
| | 128610768 | 91677669 |
| | 128618822 | 91673948 |
| | 128815316 | 91685231 |
| | 128827549 | 91687786 |

**Chr2 Base:** 12366 bp | 17951 bp | 3099 bp | 13574 bp | 2555 bp

**Chr8 Loop:** Enh D 44159 bp | Enh E 23291 bp | Enh-PC Enh G 61813 bp | MYC

**chr8:** Enh-PC Enh A,B,C 239408 bp | 12233 bp

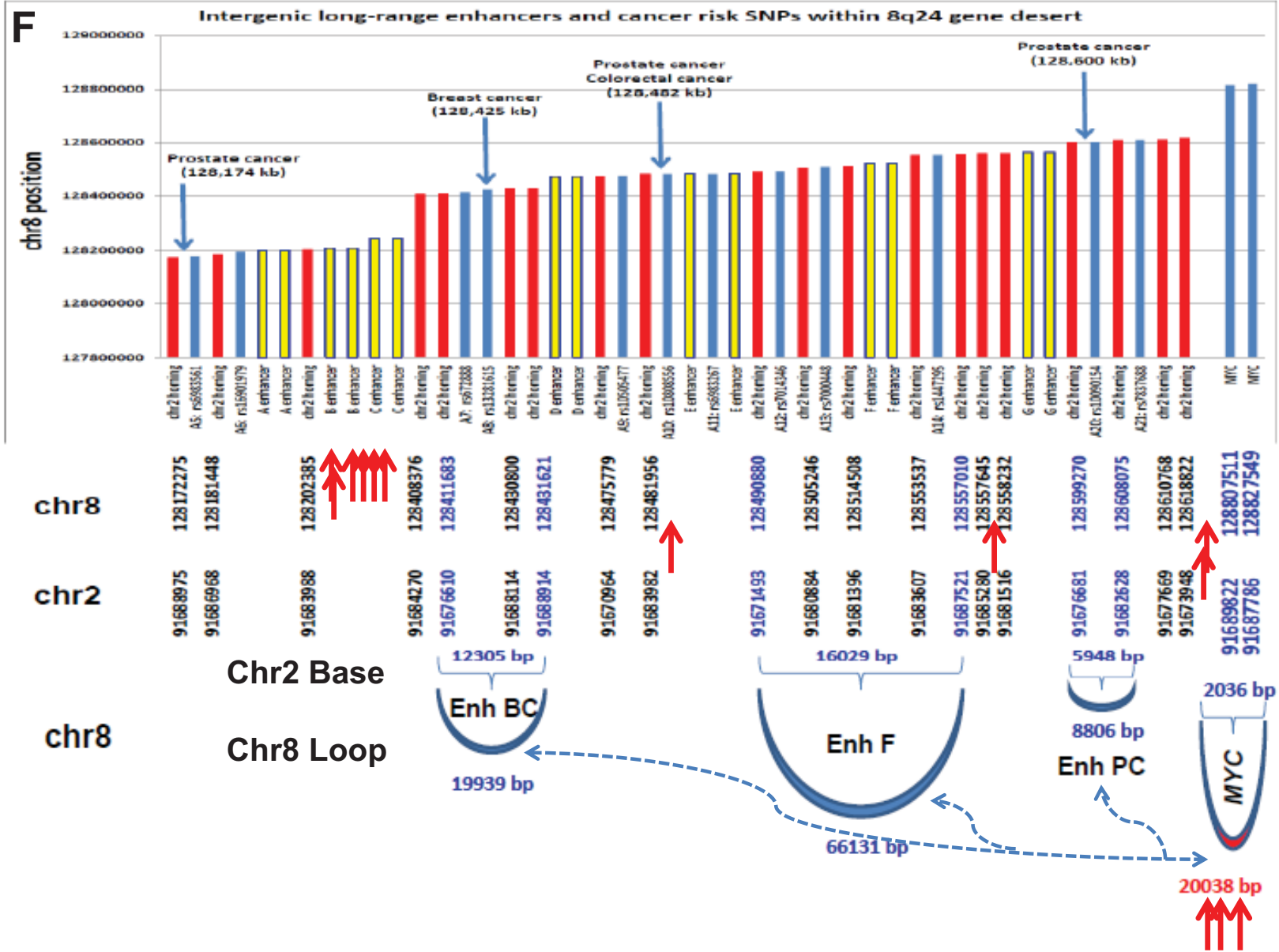

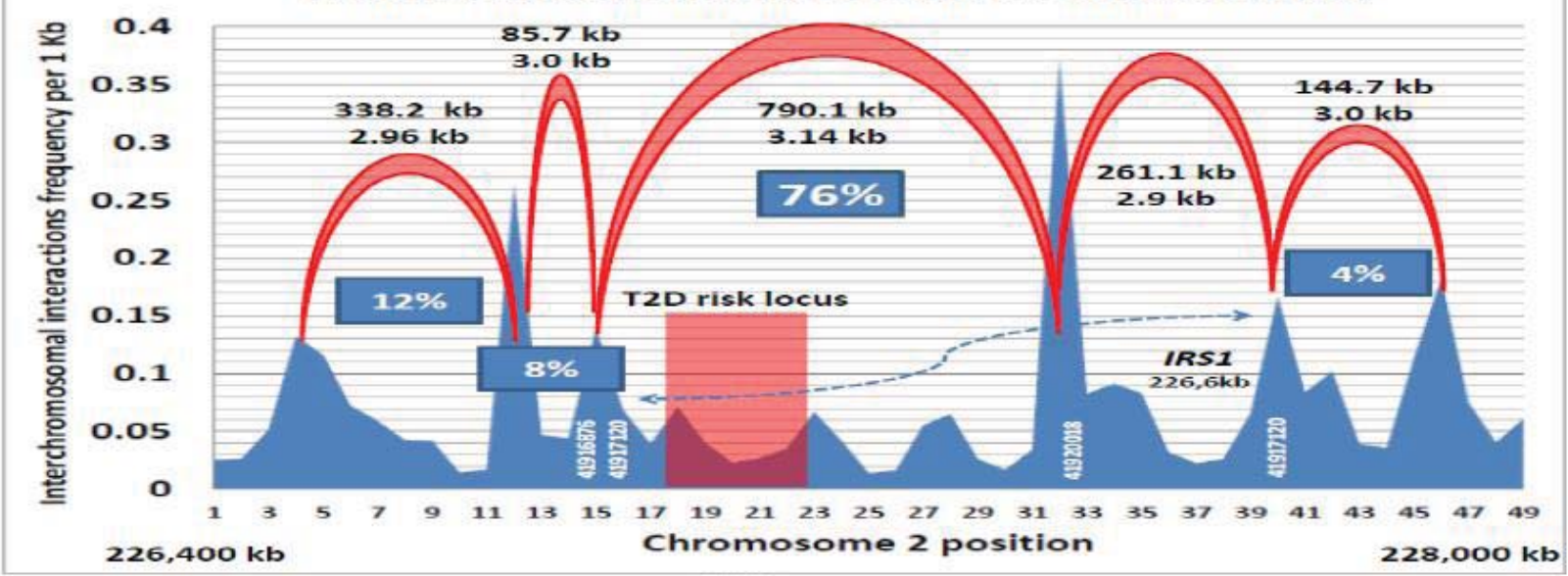
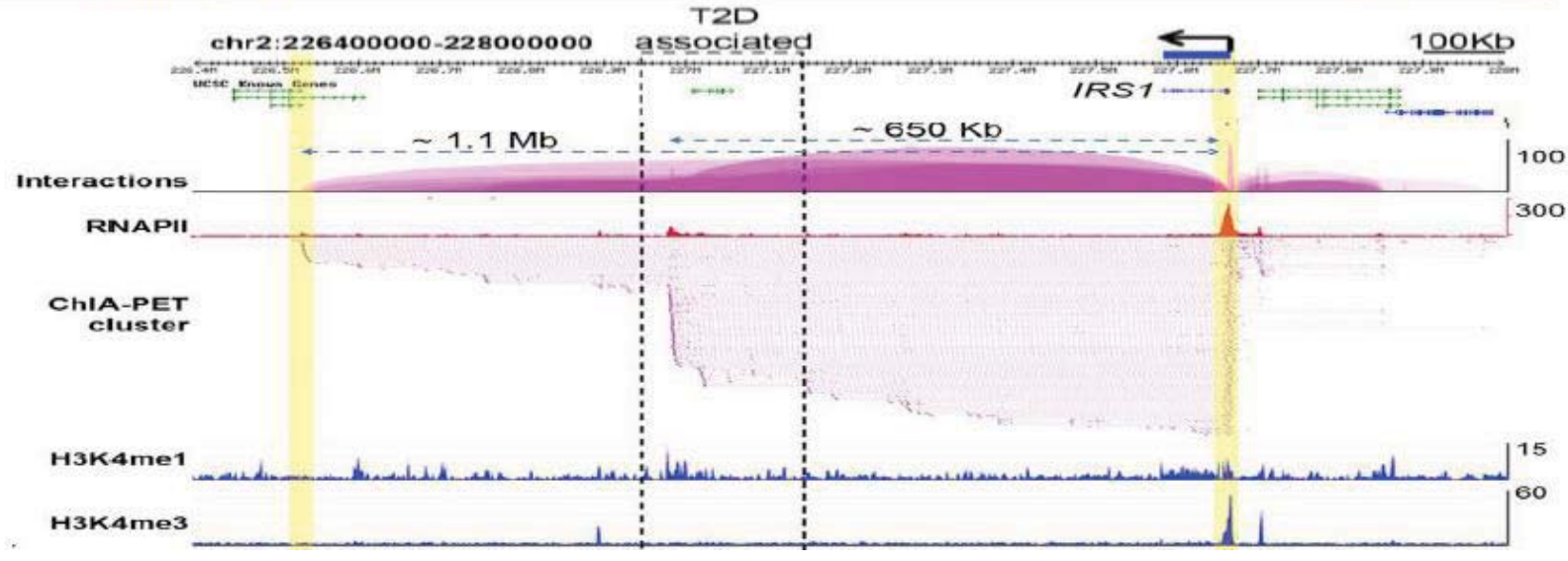

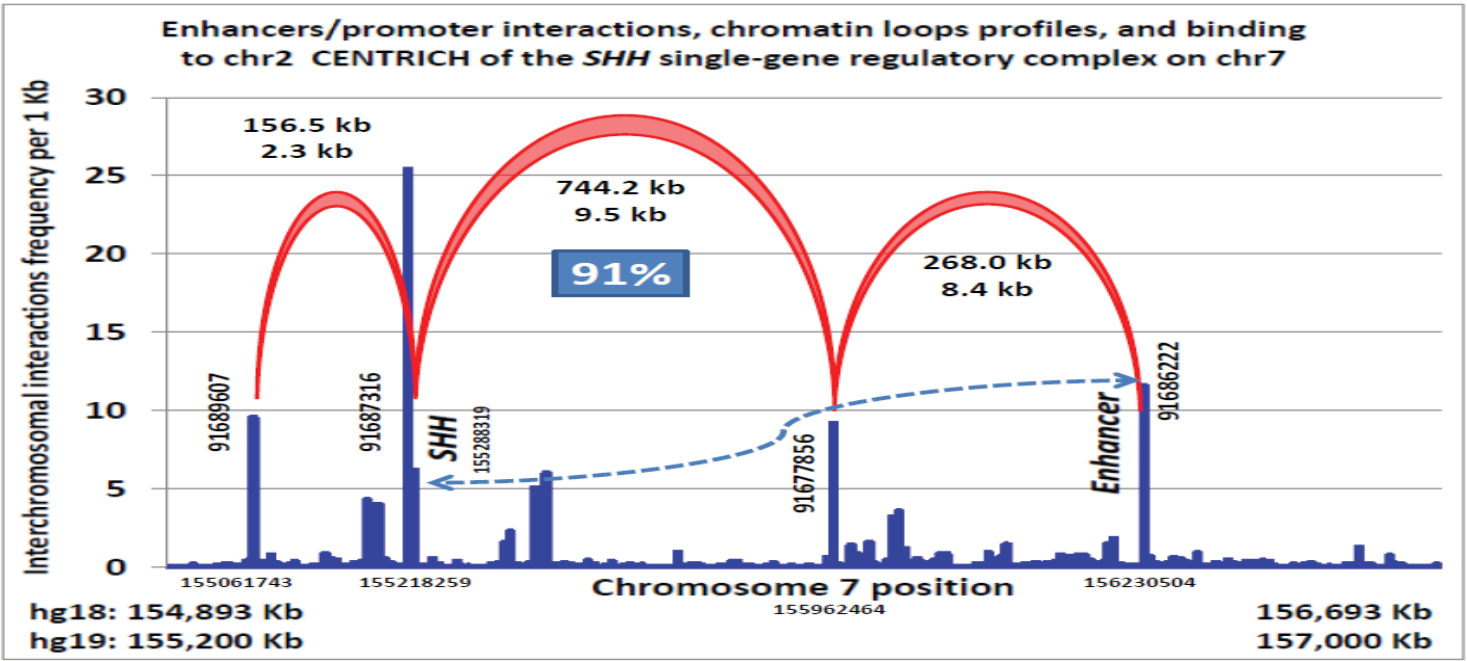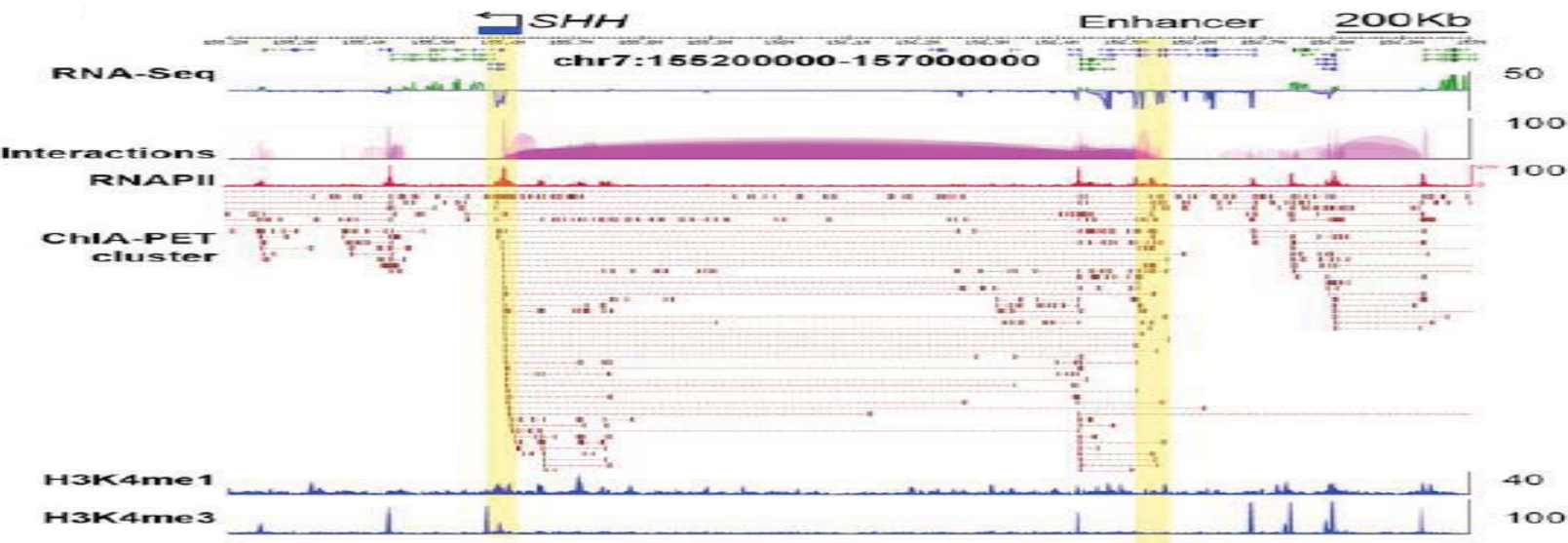

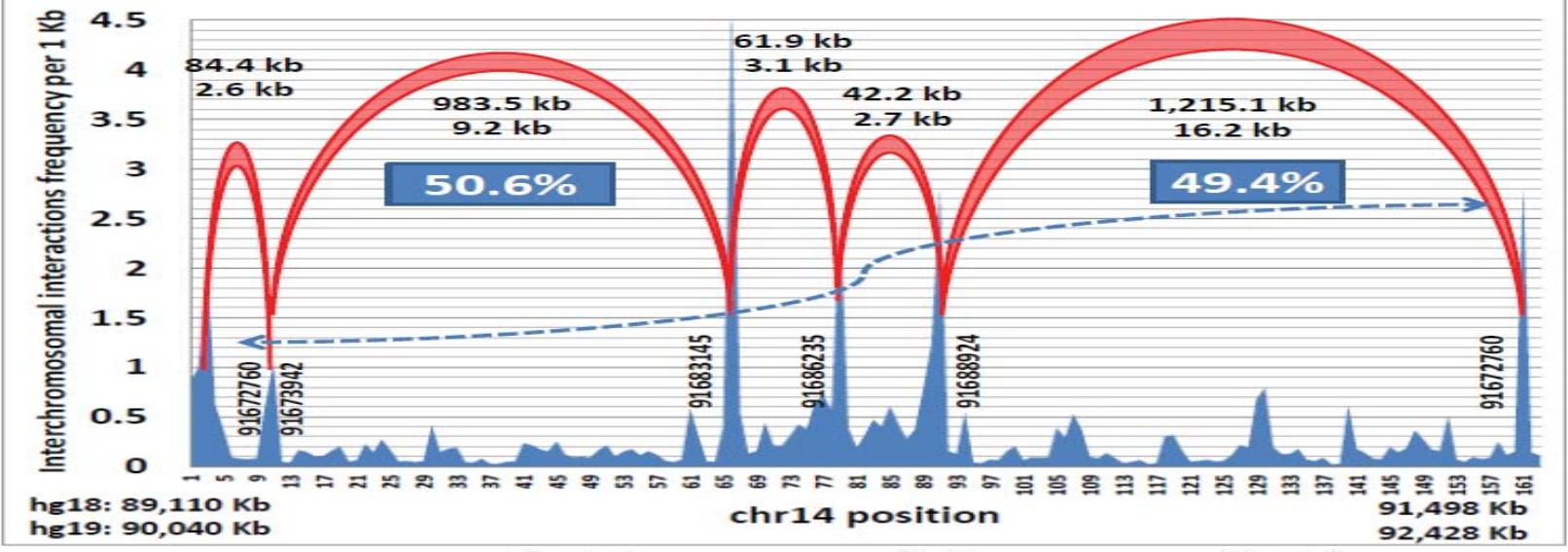
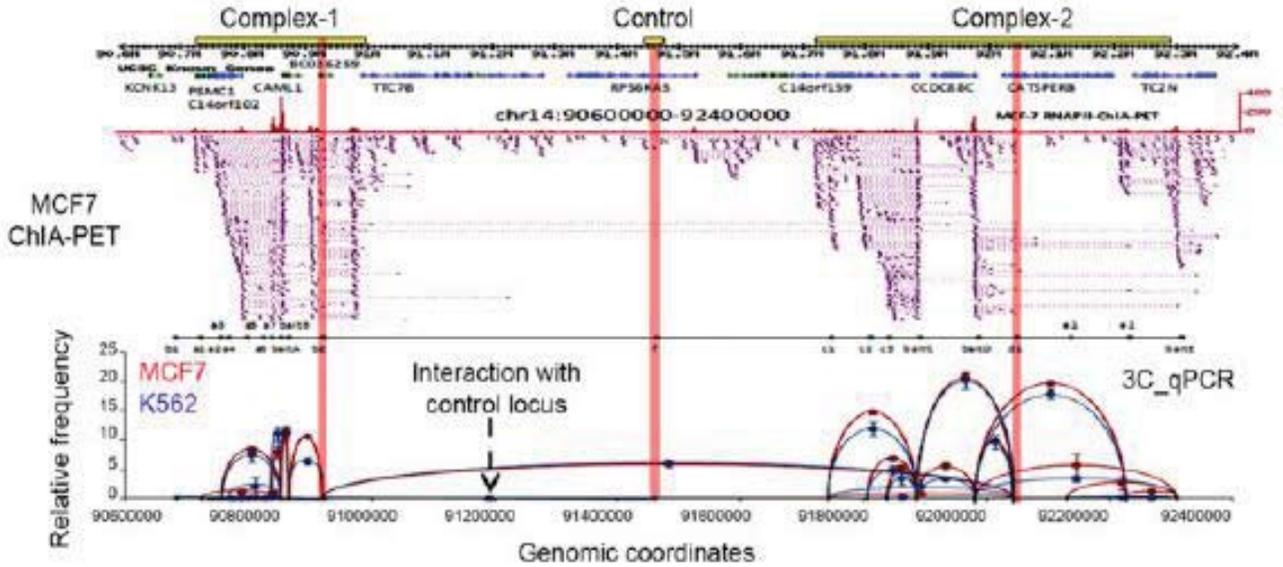

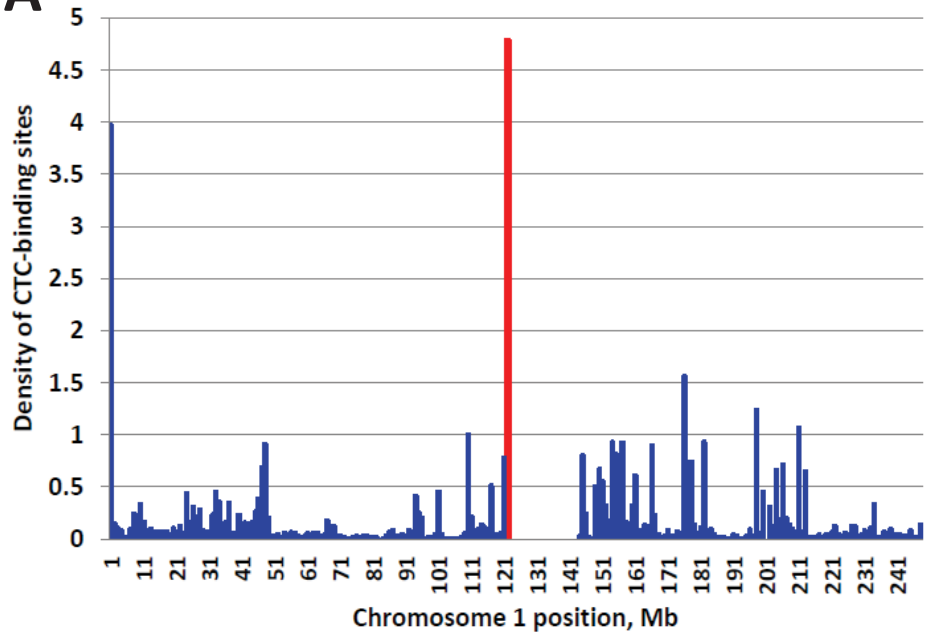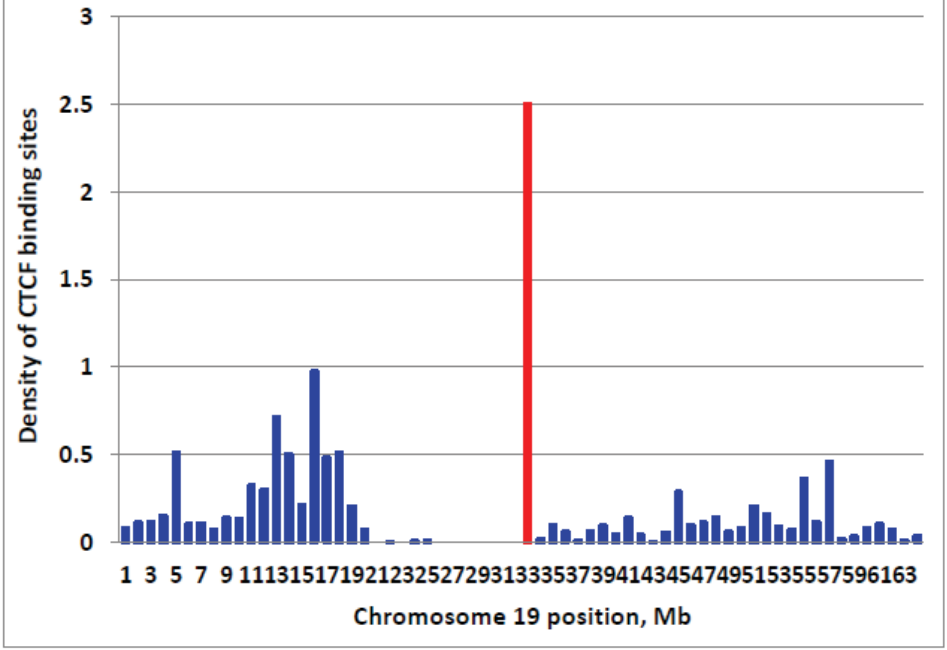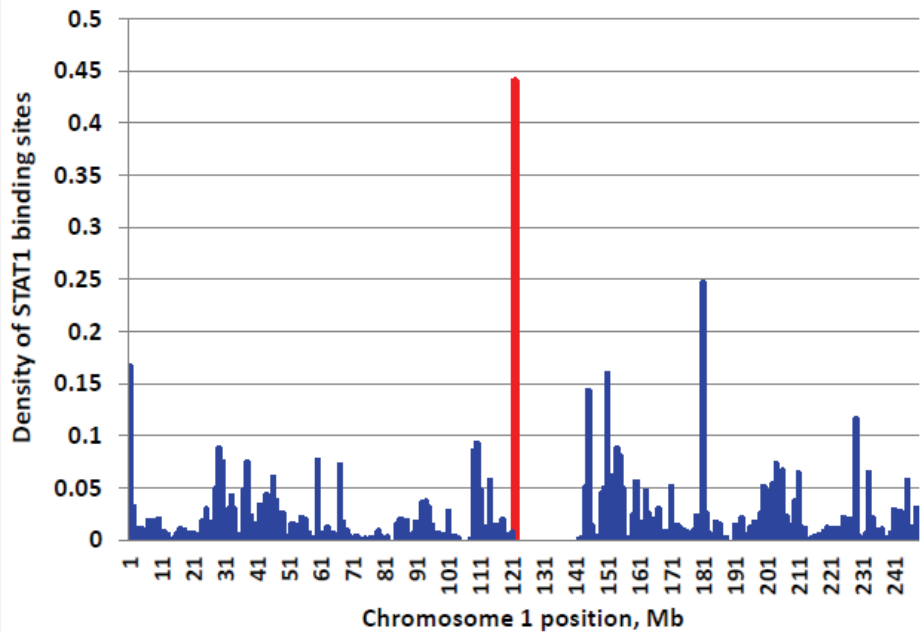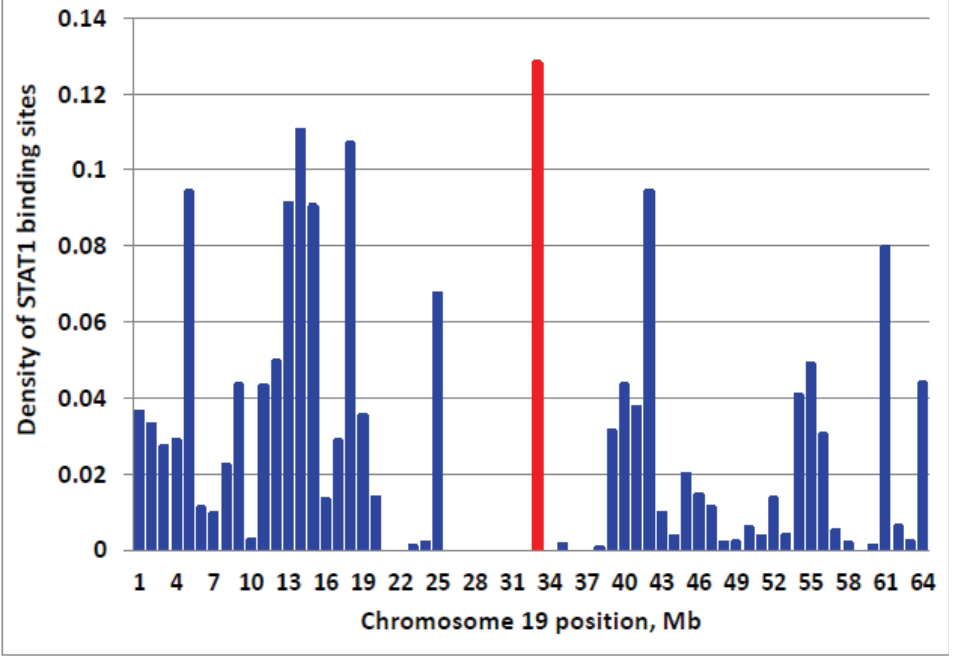

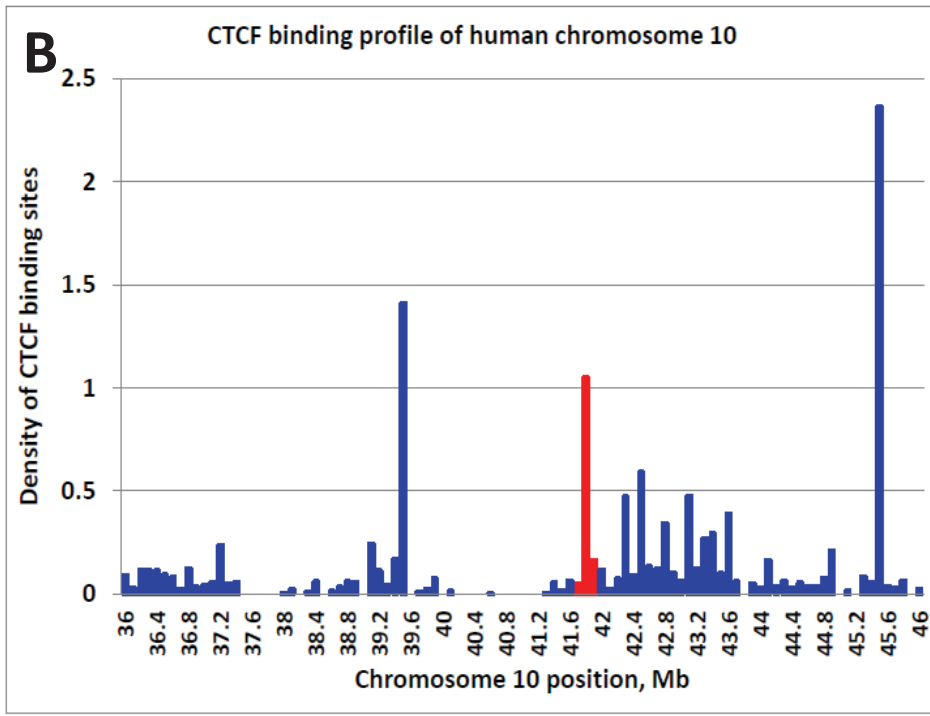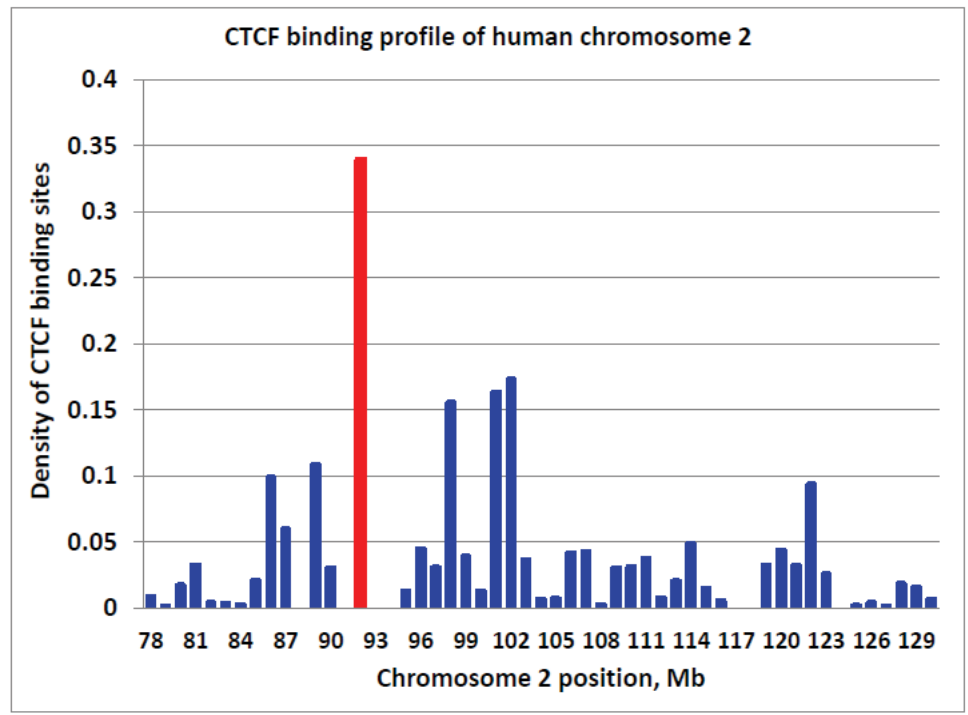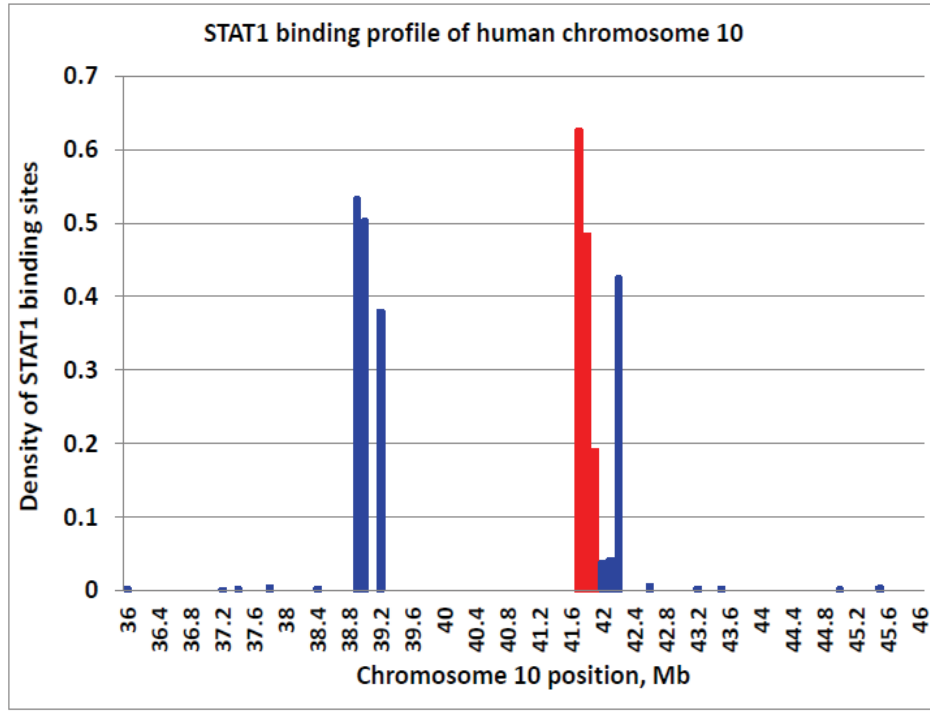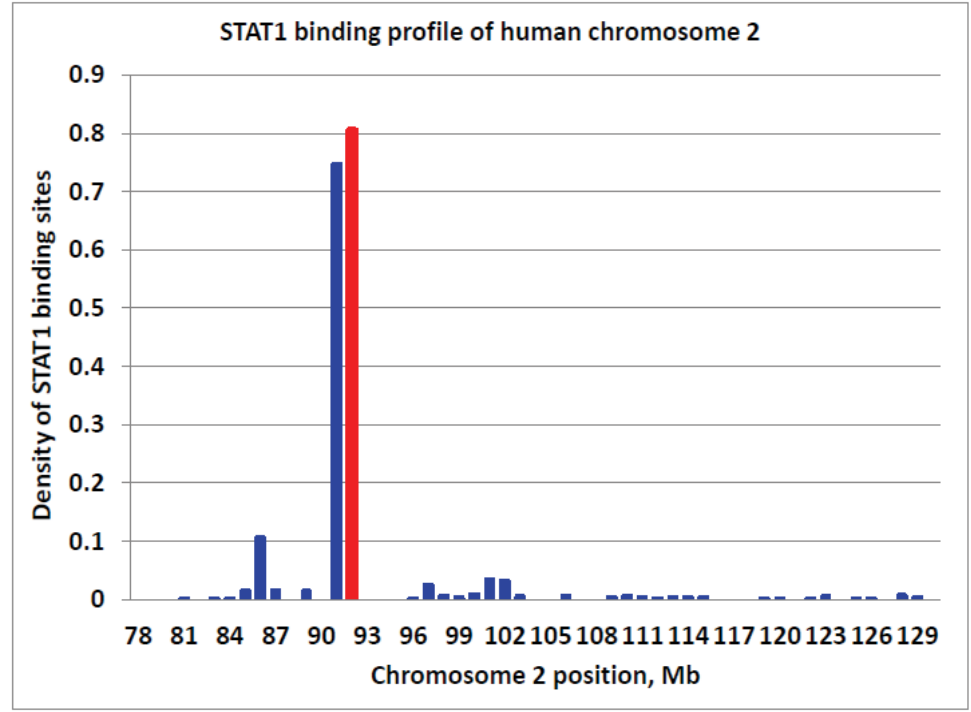

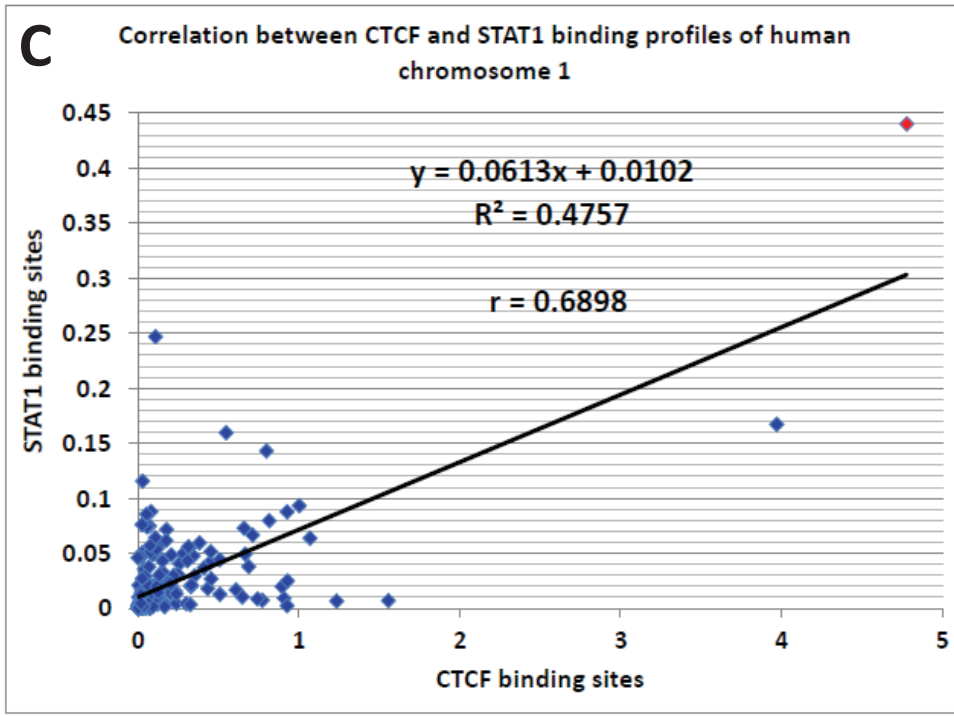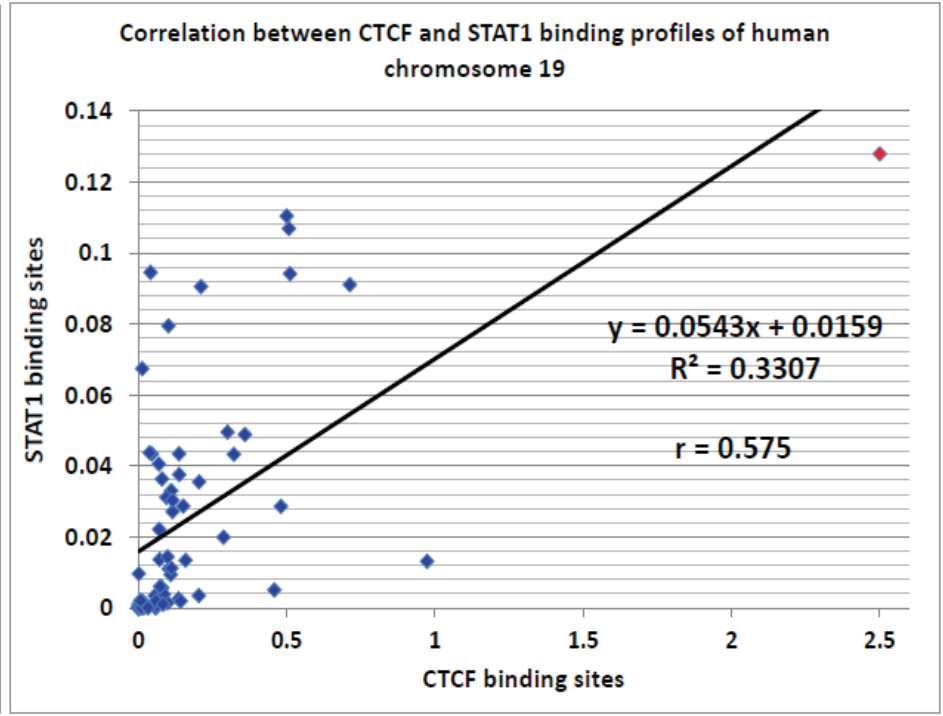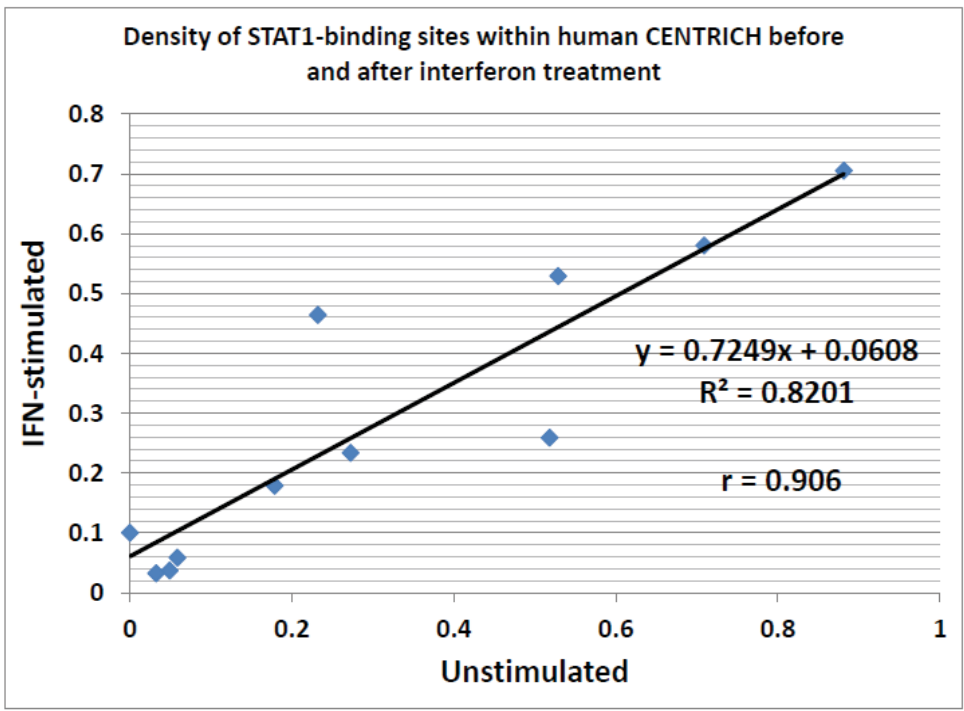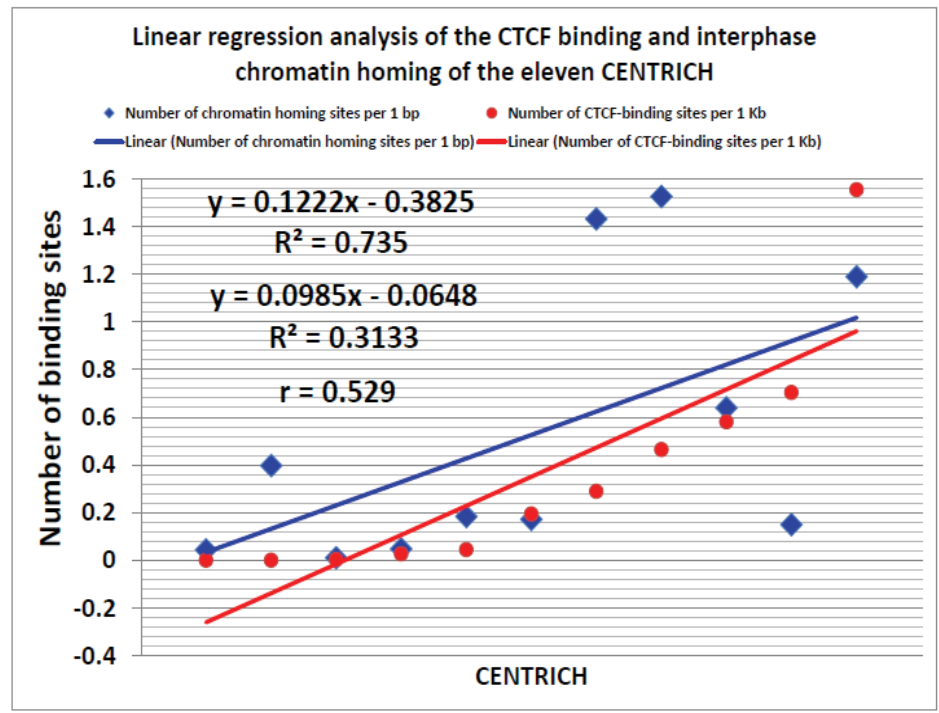

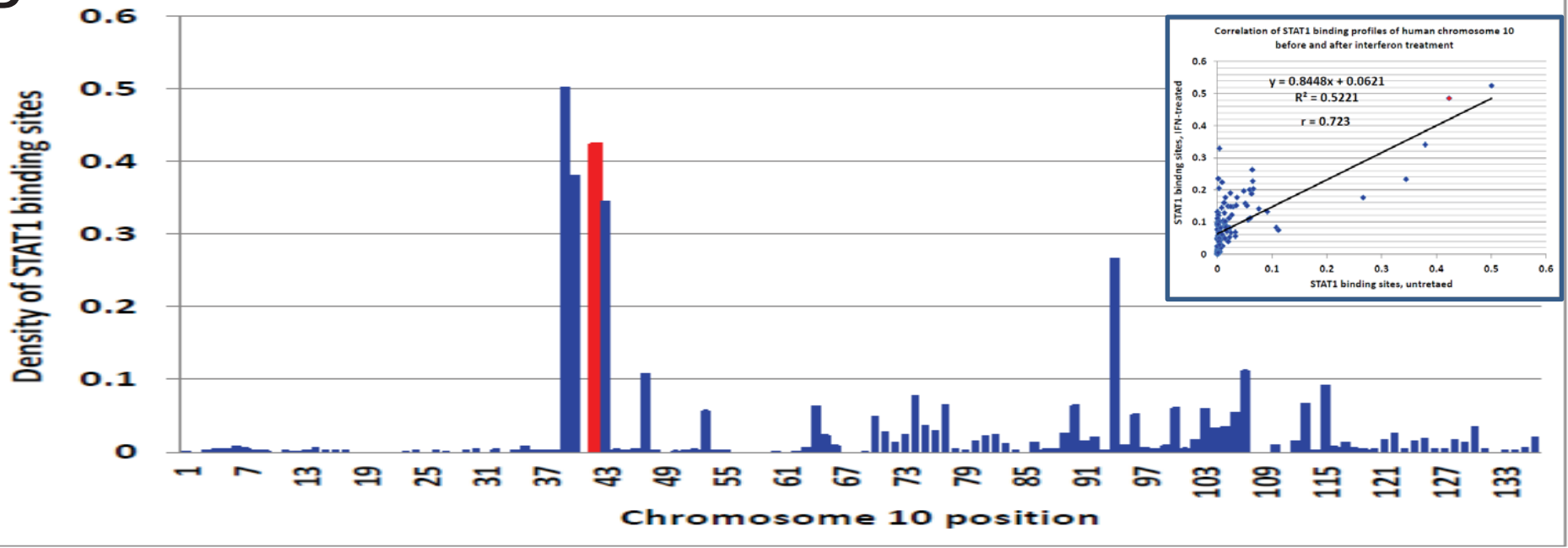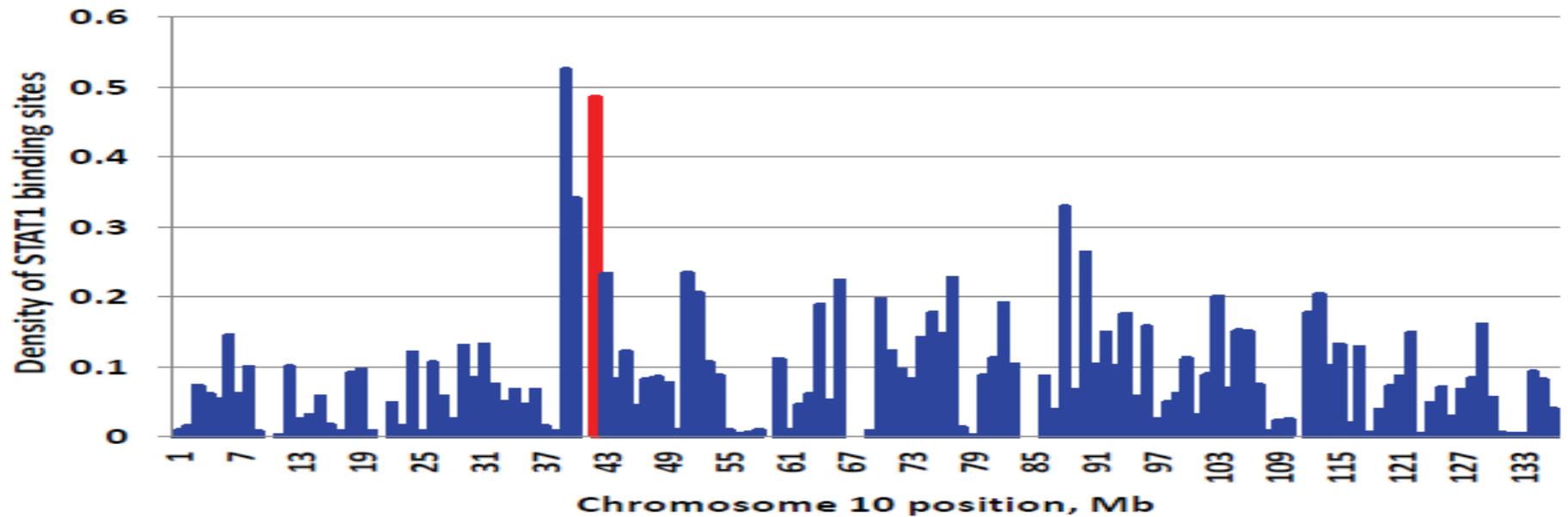

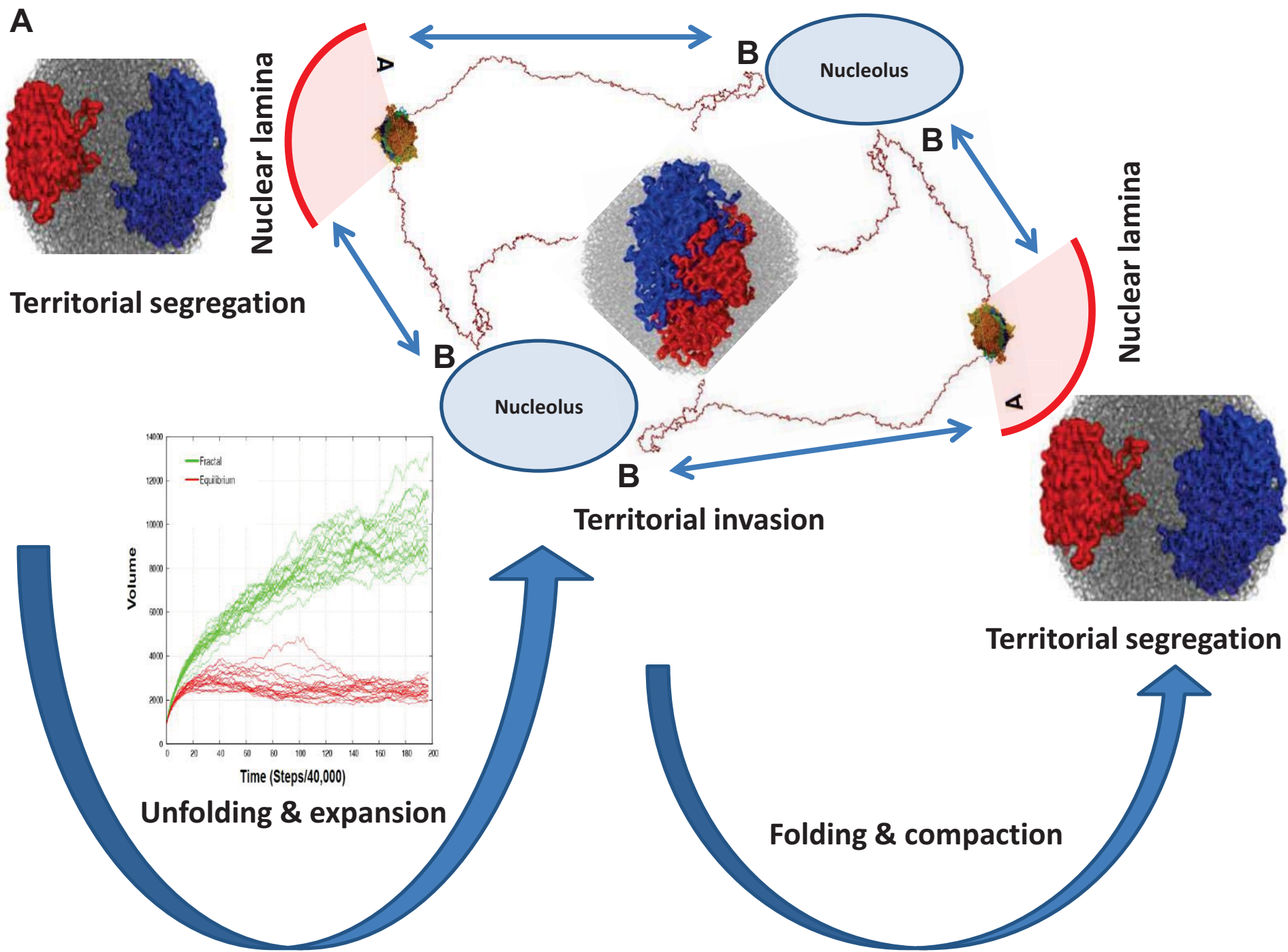

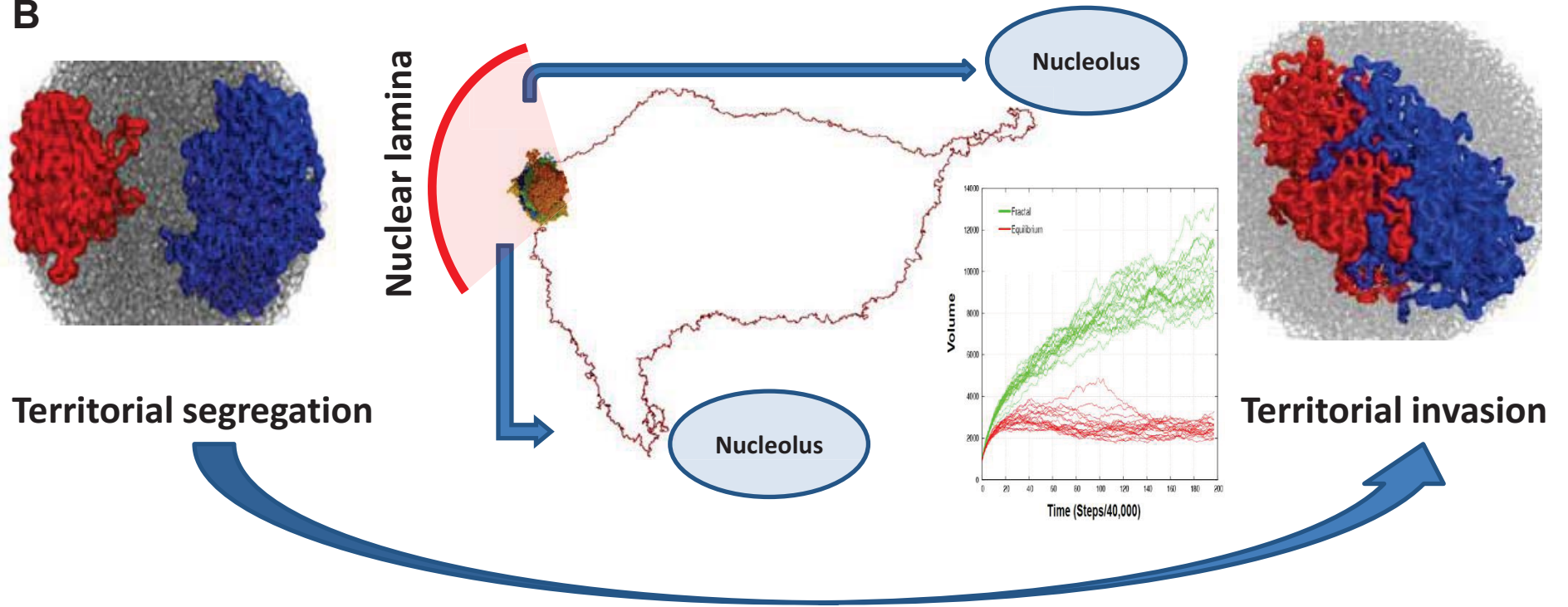